\documentclass{aa}  
\usepackage{graphicx}
\usepackage[varg]{txfonts}
\usepackage{subfig}
\usepackage{newtxtext,newtxmath}
\usepackage[labelfont=bf]{caption}

\DeclareMathOperator{\sech}{sech}
\numberwithin{equation}{section} 
\newcommand{\kms}{\,km\,s$^{-1}$}

\begin{document}

   \title{Volumetric star formation laws of disc galaxies}

   \authorrunning{C. Bacchini et al.}
   \author{Cecilia Bacchini\inst{1,2,3},
          Filippo Fraternali\inst{2,1},
          Giuliano Iorio\inst{4,1} \and
          Gabriele Pezzulli\inst{5}
          }

   \institute{Dipartimento di Fisica e Astronomia, Universit\`{a} di Bologna, via Gobetti 93/2, I-40129, Bologna, Italy\\
   \email{cecilia.bacchini@unibo.it}
   \and
   Kapteyn Astronomical Institute, University of Groningen, Landleven 12, 9747 AD Groningen, The Netherlands
   \and
   INAF - Osservatorio Astronomico di Bologna, via Gobetti 93/3, I-40129, Bologna, Italy
   \and
   Institute of Astronomy, University of Cambridge, Madingley Road, Cambridge CB3 0HA, UK
   \and
   Department of Physics, ETH Zurich, Wolfgang-Pauli-Strasse 27, 8093 Zurich, Switzerland
   }

   \date{Received October 5, 2018; accepted November 28, 2018.}

 
  \abstract{
  Star formation (SF) laws are fundamental relations between the gas content of a galaxy and its star formation rate (SFR) and 
  play key roles in galaxy evolution models. 
  In this paper, we present new empirical SF laws of disc galaxies based on volume densities.
  Following the assumption of hydrostatic equilibrium, we calculated the radial growth of the thickness of the gaseous discs in the combined gravitational 
  potential of dark matter, stars, and gas for 12 nearby star-forming galaxies. 
  This allowed us to convert the observed surface densities of gas and SFR into the deprojected volume densities. 
  We found a tight correlation with slope in the range 1.3-1.9 between the volume densities of gas (HI+H$_2$) and the SFR with a significantly smaller scatter 
  than the surface-based (Kennicutt) law and no change in the slope over five orders of magnitude. 
  This indicates that taking into account the radial increase of the thickness of galaxy discs is crucial to reconstruct their three-dimensional density profiles, 
  in particular in their outskirts. 
  Moreover, our result suggests that the break in the slope seen in the Kennicutt law is due to disc flaring rather than to a drop of the SF 
  efficiency at low surface densities. 
  Surprisingly, we discovered an unexpected correlation between the volume densities of HI and SFR, indicating that the atomic gas is a good tracer of the 
  cold star-forming gas, especially in low density HI-dominated environments.}


   \maketitle
%
\section{Introduction}\label{sec:intro}
The first formulation of an empirical star formation (SF) law was proposed by \citet{1959Schmidt} in the shape of a power law
\begin{equation}\label{eq:Slaw}
 \rho_{\mathrm{SFR}} \propto \rho_{\mathrm{HI}}^n \, ,
\end{equation}
where $\rho_{\mathrm{SFR}}$ is the star formation rate (SFR) per unit volume and $\rho_{\mathrm{HI}}$ is the HI volume density; at that time, it was not possible to 
observe molecular gas emission. 
Using the distribution of young stars in the Milky Way (MW), he suggested that the index $n$ of this power law is between 2 and 3. 
Unfortunately, if we focus on galaxies outside the MW, we can directly observe only the projected quantities, for example the surface densities, so  Schmidt's 
approach is less suitable. 

The works of \citet{1989Kennicutt,1998Kennicutt} set the current standard method to investigate SF law. 
Using a sample of nearby star-forming galaxies, \citet{1989Kennicutt} derived a relation involving the radial profiles of the gas and the SFR surface densities 
\citep[see also][]{2007Kennicutt,2001MartinKennicutt}.  
The so-called Kennicutt (or Schmidt-Kennicutt) law is
\begin{equation}\label{eq:Klaw}
 \Sigma_\mathrm{SFR} \propto \Sigma_\mathrm{gas} ^ N \, ,
\end{equation}
where $\Sigma_{\mathrm{SFR}}$ and $\Sigma_{\mathrm{gas}}$ are the surface densities of SFR and total gas (HI+H$_2$). 
However, this surface-based power law showed a break at densities below a threshold value. 
Later, \cite{1998Kennicutt} collected a sample of spiral galaxies and starbursts to study the SF law over a range of seven orders of magnitude. 
Using surface densities integrated over the entire disc, he found a single power-law correlation with an index of $N = 1.4 \pm 0.15$. 
In the next two decades, there was much work on two main issues of the SF law: the first issue refers to the 
gas phase that better correlates with SF and the second concerns the possibility that the power-law index changes in particular environments, producing a break in the 
relation. 

About the first issue, several observational studies claimed that the vital fuel of SF is molecular gas. 
Indeed, a gas cloud can gravitationally collapse only if its temperature is low enough, and molecules are very efficient coolants. 
\cite{2008Bigiel} studied the SFR-H$_2$ relation in 18 nearby galaxies through pixel-to-pixel analysis and radial profiles extraction. 
These authors found, on a sub-kiloparsec scale, a linear correlation between SFR and molecular gas surface densities 
\citep[see also][]{2002Wong,2007Kennicutt,2011Bolatto,2011Schruba,2012Marasco,2013Leroy}.
Similarly, \cite{2010Lada} found a linear relation linking the mass of Galactic molecular clouds and the number of hosted young stellar objects. 

However, molecular clouds form from atomic gas and, after the SF has occurred, they are destroyed by stellar feedback. 
Hence, we would expect the atomic or total gas to correlate with SFR, as originally found by \cite{1959Schmidt}. 
On the contrary, \cite{2008Leroy} found no correlation on sub-kiloparsec scale between HI and SFR in nearby star-forming galaxies \citep[see also][]{2007Kennicutt}. 
The picture changes in HI-dominated environments such as the outskirts of spiral galaxies and dwarf galaxies, where the SFR seems to correlate also with 
atomic gas, but the efficiency of SF drammatically drops at these low surface densities \citep[e.g.][]{1998Ferguson,2010Bigiel,2011Bolatto,2011Schruba,2016Yim}. 

The second issue about the Kennicutt law concerns the shape of the classical relation involving the total gas and the SFR. 
\cite{2008Bigiel} found that the Kennicutt law index changes at $\Sigma_\mathrm{gas} \approx 9 \, \mathrm{M}_\odot \mathrm{pc}^{-2}$, which approximately 
corresponds to the transition to low density and HI-dominanted environments \citep[see also][]{2011Bolatto,2011Schruba,2014Dessauges}. 
As a consequence, a surface-based, double power-law relation was proposed and its break was explained as an abrupt change in the efficiency of SF at 
a specific threshold density \citep[see also][]{2004Schaye}. 
Similarly, \cite{2015Roychowdhury} studied the Kennicutt law in HI-dominanted regions of nearby spirals and dwarf irregulars, and found a power-law relation with slope 1.5 for both kind of galaxies. 
Their relation showed however an offset of one order of magnitude with respect to the SF law for more central regions, where the gas surface density is higher. 
Despite that the existence of the break is not firmly confirmed as several authors recovered the classical single power law, sometimes with a different 
index. 
For example, \cite{2003Boissier} and \cite{2012Barnes} estimated $N \approx 2$ and $N=2.8 \pm 0.3 $, respectively, in nearby star-forming galaxies, 
\cite{2004Heyer} found $N \approx 3.3$ for M33, and \cite{2017Sofue} measured $N=1.12 \pm 0.37$ in the MW. 

From a theoretical point of view, it is possible to predict the power-law index assuming that a given physical process regulates the birth of stars. 
The simplest model involves the gravitational collapse and the SF timescale is set by the free-fall time \citep{1977Madore}.
As a result, the SFR is given by the fraction of gas converted into stars per free-fall time, so $\rho_\mathrm{SFR} \propto \rho_\mathrm{gas}^{1.5}$. 
This corresponds to $\Sigma_\mathrm{SFR} \propto \Sigma_\mathrm{gas}^{1.5}$ if the disc thickness is constant with the galactocentric radius. 
This basic model has been proposed to explain the observed Kennicutt law with index $N \approx 1.4$. 
Another possible SF timescale is the orbital time in the disc, which is related to the rotation velocity of the galaxy 
\citep[e.g.][]{1998Kennicutt,2007Kennicutt,2003Boissier,2017Bolatto}. 
Alternatively, if the balance between turbulent motions and gravity is assumed to regulate SF, the predicted slope is $N \approx 2$ 
\citep[e.g.][]{1981Larson,2015Elmegreen}. 
Another class of models aims to predict the critical density for the broken power law. 
For example, \cite{1964Toomre} formalism allows us to estimate the critical density above which a gas disc is gravitationally unstable and 
the shear is low enough to have SF \citep[e.g.][]{1989Kennicutt,1990PhDTRomeo,1992Romeo,1998aHunter,2001MartinKennicutt}. 
These are only a few examples among the plenty of possible models that have been proposed to explain or predict observations 
(see \citealt{2014Krumholz_rev} and references within). 

Overall, the picture is very complex and the shape of the relation between gas and SF remains unknown. 
Moreover, it is unclear which gas phase matters most for SF, whether molecular or atomic or both. 
Having a robust recipe for SF is very important, as the SF law is a key ingredient of numerical simulations and theoretical models of galaxies formation and 
evolution (e.g. chemical evolution of discs). 
All the SF laws mentioned so far \citep[except][]{1959Schmidt} are based on surface densities because they are easy to observe, but the volume densities 
are likely more physically meaningful quantities than surface densities. 
In addition, gas discs in galaxies are expected to be nearly in hydrostatic equilibrium, so their thickness grows going from the inner radii to the outskirts
and the resulting projection effects are not negligible \citep[e.g.][]{1981VanderKruit,2008Abramova,2011Banerjee,2015Elmegreen}. 
The purpose of this paper is to build a volumetric star formation (VSF) law through a method to convert surface densities to volume densities 
in local disc galaxies. 
The general model is described in Sec.~\ref{sec:model}, then Sec.~\ref{sec:sample} explains how we selected the sample of galaxies to test this model. 
The thickness of the gas disc for each galaxy is calculated in Sec.~\ref{sec:thick} and the resulting volumetric correlations are shown in Sec.~\ref{sec:vsf}. 
In Sec.~\ref{sec:discussion}, our results are discussed and compared to other works. 
Finally, we provide summary and conclusions in Sec.~\ref{sec:concl}. 

\section{Volume densities from hydrostatic equilibrium}\label{sec:model}
In order to build the VSF law, we need the volume densities of atomic gas ($\rho_\mathrm{HI}$), molecular gas ($\rho_\mathrm{H_2}$), and SFR 
($\rho_\mathrm{SFR}$). 
In the following, we show how the simple assumption of the vertical hydrostatic equilibrium allows us to estimate these quantities and the ingredients 
that are needed to calculate them. 

\subsection{Hydrostatic equilibrium}
Let us consider a rotating disc of gas in hydrostatic equilibrium in the gravitational potential $\Phi$ of a galaxy, which is assumed to be symmetric with 
respect to the rotation axis (axisymmetry) and the plane $z=0$ (midplane). 
The vertical distribution of the gas density $\rho(R,z)$ can be described by the stationary Euler equation in the $z$ direction as follows:
\begin{equation}\label{eq:euler}
\frac{\partial \Phi(R,z)}{\partial z} = - \frac{1}{\rho(R,z)} \frac{\partial P(R,z)}{\partial z} \, ,
\end{equation}
where $P(R,z)$ is the gas pressure due to the combination of thermal and turbulent motions, the latter being the dominant component. 
At a given galactocentric radius $R$, we take the three components of the velocity dispersion of the gas to have the same value in all directions 
$\sigma_x(R) = \sigma_y(R) = \sigma_z(R) = \sigma(R)$ (isotropy). 
Then, we assume that the velocity dispersion $\sigma$ is constant along $z$ (vertically isothermal gas). 
Therefore, on galactic scales, the global profile of $\sigma$ depends only on $R$ and the gas pressure can be written as \citep[e.g.][]{1995Olling}
\begin{equation}
  P(R,z) = \sigma^2(R) \rho(R,z)
\end{equation}
and Eq.~\ref{eq:euler} can be solved for the density profile 
\begin{equation}\label{eq:densprof}
\rho(R,z) = \rho(R,0) \exp \left[- \frac{\Phi(R,z) - \Phi(R,0)}{\sigma^2(R)} \right] \, ,
\end{equation}
where $\rho(R,0)$ and $\Phi(R,0)$ are the radial profiles of the gas volume density and the total gravitational potential evaluated in the midplane of 
the galaxy. 

\subsection{Gravitational potential}\label{sec:pot_model}
The gravitational potential of a galaxy can be obtained through the Poisson equation for gravity once its mass distribution is known. 
The main mass components of star-forming galaxies are dark matter (DM), stars in the form of a disc and a bulge (if present), and gas. 

\subsubsection{Dark matter halo}\label{sec:dm_model}
The DM distribution can be modelled as a pseudo-isothermal halo \citep{1985Vanalbada} or a Navarro-Frenk-White (NFW) halo \citep{1996NavarroFrenkWhite}. 
For simplicity, the DM halo distribution is assumed spherical.  
The pseudo-isothermal density profile is
\begin{equation}\label{eq:isohalo}
 \rho_\mathrm{DM}(r) = \rho_\mathrm{DM,0} \left( 1 + \frac{r^2}{r_\mathrm{c}^2} \right)^{-1} \, ,
\end{equation}
where $\rho_{\mathrm{DM},0}$ is the central volume density and $r_\mathrm{c}$ the core radius. 
The NFW profile is
\begin{equation}\label{eq:nfwhalo}
 \rho_\mathrm{DM}(r) = \rho_\mathrm{DM,0} \left(\frac{r}{r_\mathrm{s}} \right)^{-1} \left( 1 + \frac{r}{r_\mathrm{s}} \right)^{-2}\, ,
\end{equation}
where $c=r_{200}/r_\mathrm{s}$ is the concentration parameter; $r_{200}$ is the radius within which the average density contrast with respect to the 
critical density of the Universe equals 200. 
The spherical radius is $r=\sqrt{R^2+z^2}$ in cylindrical coordinates. 

\subsubsection{Stellar disc}\label{sec:disc_model}
The stellar disc mass distribution is modelled with an exponential radial profile and a $\sech^2$ vertical profile \citep{1981VanderKruit_a},
\begin{equation}\label{eq:expsec_stardisc}
 \rho_\star (R,z) = \rho_{\star,0} \exp \left( - \frac{R}{R_\star} \right) \sech^2 \left(\frac{z}{z_\star} \right) \, ,
\end{equation}
where $\rho_{\star,0}$ is the central density, $R_\star$ is the stellar scale length, and $z_\star$ the scale height, which is assumed to be 
$z_\star=R_\star/5$ (see \citealt{2011vanderKruitFreeman} and references within). 

\subsubsection{Stellar bulge}\label{sec:bulge_model}
The bulge mass distribution is modelled using a sphere with exponential profile,
\begin{equation}\label{eq:bulge_expball}
 \rho_\mathrm{b}(r) = \rho_{\mathrm{b},0} \exp \left( - \frac{r}{r_\mathrm{b}} \right) \, ,
\end{equation}
where $\rho_{\mathrm{b},0}$ and $r_\mathrm{b}$ are central density and scale radius. 
The justification for the choice of Eq.~\ref{eq:bulge_expball} is discussed in Sec.~\ref{sec:bulge}. 

\subsubsection{Gas surface density}\label{sec:gas_model}
In order to model the variety of gas distributions in galaxies (both for the atomic and molecular phases), we need a flexible model. 
Hence, we combined a polynomial and an exponential function\begin{equation}\label{eq:poly}
  \Sigma (R) = \Sigma_0 \left( 1 + C_1 R + C_2 R^2 + C_3 R^3 + C_4 R^4 \right) \exp \left( -\frac{R}{R_\Sigma} \right) \,,
\end{equation}
where $\Sigma_0$ is the central surface density, $R_\Sigma$ is the scale radius, and $C_i$ are the polynomial coefficients. 

\subsection{Velocity dispersion}\label{sec:disp_model}
In previous works \citep[e.g.][]{2008Abramova,2008Leroy,2015Elmegreen}, the gas velocity dispersion was assumed to be constant with radius. 
On the contrary, several measurements of the velocity dispersion in nearby galaxies and in the MW show that it decreases with increasing 
galactocentric radius, following an exponential or linear trend \citep[e.g.][]{2002Fraternali,2008Boomsma,2009Tamburro,2016Mogotsi,2017Marasco}. 
Hence, we derived the profile of $\sigma(R)$ from the observations (Sec.~\ref{sec:hivdisp}) and modelled it, for the atomic and molecular 
phases, with the exponential function 
\begin{equation}\label{eq:exp_vdisp}
 \sigma(R) = \sigma_{0} \exp \left( -\frac{R}{R_\sigma} \right) \, ,
\end{equation}
where $\sigma_{0}$  is the velocity dispersion at the galaxy centre and $R_\sigma$ is a scale radius. 
This function can also adequately model a linear decline for large $R_\sigma$ compared to the galaxy size. 

\subsection{Scale height definition}\label{sec:hdefapprox}
By means of a second order Taylor expansion of $\Phi$ \citep[see e.g.][]{1995Olling,2009Koyama}, Eq.~\ref{eq:densprof} can be approximated near the midplane by a 
Gaussian profile, 
\begin{equation}\label{eq:densprofscale}
\rho(R,z) =  \rho (R,0) \exp \left[ - \frac{z^2}{2 h^2 (R)} \right] \, ,
\end{equation}
where the radial profile of the vertical scale height $h(R)$ is
\begin{equation}\label{eq:hdef}
 h(R) \equiv \sigma(R) \left[ \frac{\partial^2 \Phi(R,0)}{\partial z^2} \right]^{-\frac{1}{2}} \, .
\end{equation}
The roles of the gravitational potential and the velocity dispersion are opposite, as the first drags the gas towards the midplane, while the second gives 
rise to a force directed upward. 
As shown in Sec.~\ref{sec:thick}, in real galaxies both terms decrease with radius, but the global result is an increase of the scale height 
with radius. 

Eq.~\ref{eq:hdef} is an analytical approximation for the scale height and it is valid if the vertical gradient of the gravitational 
potential is null within small heights above the midplane. 
In addition, Eq.~\ref{eq:hdef} does not take into account the self-gravity of the gas, which could become significant at large radii. 
As a consequence, we do not calculate the scale height analytically with Eq.~\ref{eq:hdef}, but we use a numerical method to estimate the scale height from 
Eq.~\ref{eq:densprof}. 
In Appendix~\ref{ap:h_analyt}, we however show that this approximation is not as coarse as it may seem, but it gives results that are compatible with 
the numerical scale heights. 

\subsection{From surface densities to volume densities}
Let us now look at the gas disc from the perspective of an external observer who measures the radial profile of the gas density; we are assuming a 
face-on disc for simplicity. 
The observed profile of the surface density is the projection along the line of sight of the corresponding volume density profile as follows:
\begin{equation}\label{eq:project}
  \Sigma(R) = 2 \int_{0}^{+\infty}\rho(R,z) dz \, .
\end{equation}
Substituting Eq.~\ref{eq:densprofscale} in Eq.~\ref{eq:project} and solving the integral, we obtain the volume density in the midplane 
\begin{equation}\label{eq:midplane}
 \rho(R,0) = \frac{\Sigma(R)}{\sqrt{2 \pi} h(R)} \, .
\end{equation}
Hence, Eq.~\ref{eq:midplane} gives us the volume density from the observed surface density and the scale height. 
This is valid for any component, in particular HI, H$_2$ and SFR. 
The gaseous and the SFR components require separate brief discussions. 

\subsubsection{Gas volume densities}\label{sec:gas_vol_dens}
As mentioned in Sec.~\ref{sec:hdefapprox}, the scale height of a gas disc depends on the velocity dispersion of the gas. 
The molecular and the atomic phase are characterised by different values for the velocity dispersion \citep[e.g.][]{2016Mogotsi,2017Marasco}. 
Hence, we must consider these components as distributed into two separate discs both in hydrostatic equilibrium and each one with its own scale height 
($h_\mathrm{HI}$ and $h_\mathrm{H_2}$). 
Therefore, Eq.~\ref{eq:midplane} can be written both for HI and H$_2$ and the volume density of the total gas (HI+H$_2$) in the midplane becomes
\begin{equation}
\begin{split}
 \rho_\mathrm{gas}(R,0) & = \rho_\mathrm{HI}(R,0) + \rho_\mathrm{H_2}(R,0) =\\
  & = \frac{\Sigma_\mathrm{HI}(R)}{\sqrt{2 \pi} h_\mathrm{HI}(R)} + \frac{\Sigma_\mathrm{H_2}(R)}{\sqrt{2 \pi} h_\mathrm{H_2}(R)} \, .
\end{split}
\end{equation}
In this way we defined three quantities ($\rho_\mathrm{HI}$, $\rho_\mathrm{H_2}$ and $\rho_\mathrm{gas}$) that we compare to the SFR 
volume density. 

\subsubsection{Star formation rate volume density}\label{sec:sfr_vol_dens}
The SFR vertical distribution is not known a priori but, as stars form from gas, it is reasonable to assume that an equation analogous to 
Eq.~\ref{eq:midplane} applies to newborn stars as well, given some suitable definition of the SFR scale height ($h_\mathrm{SFR}$). 
For this latter, we decided to make two extreme assumptions. 
The first consists in supposing that $h_\mathrm{SFR}$ is a function of the scale heights of the two gas phases. 
Thus, we assumed it to be the mean of the scale heights of both gas phases weighted for the respective gas fractions,
\begin{equation}\label{eq:hSFR_wmean}
 h_\mathrm{SFR}(R) = h_\mathrm{HI}(R) f_\mathrm{HI}(R) + h_{\mathrm{H}_2}(R) f_{\mathrm{H}_2}(R) \, ,
\end{equation}
where $f_\mathrm{HI}(R)=\Sigma_\mathrm{HI}(R)/\Sigma_\mathrm{gas}(R)$ and $f_{\mathrm{H}_2}(R)=\Sigma_\mathrm{H_2}(R)/\Sigma_\mathrm{gas}(R)$ are the 
fraction of HI and H$_2$ with respect to the total gas. 
With this choice, if the atomic gas is fully dominant with respect to the molecular phase (as in the outskirts of spirals and in dwarfs), $h_\mathrm{SFR}(R)$ 
coincides with $h_\mathrm{HI}$ and viceversa with $h_\mathrm{H_2}$. 
If both gas phases are present in a comparable amount, then $h_\mathrm{SFR}(R)$ is simply a weighted mean of $h_{\mathrm{H}_2}$ and $h_\mathrm{HI}$. 
For the second choice, we assumed a constant $h_\mathrm{SFR}$, we took $h_\mathrm{SFR}=100$ pc as a fiducial value \citep{2012Barnes}. 
We note that choosing a different constant would change only the normalisation factor for the SFR volume density. 
It is reasonable to expect that the true SFR scale height lies between these two extreme choices. 
We could also consider $h_\mathrm{HI}$ or $h_{\mathrm{H}_2}$ as alternative definitions of $h_\mathrm{SFR}(R)$. 
We explore these cases in Sec.~\ref{sec:hionly_corr} and  Sec.~\ref{sec:h2only_corr}. 

%
\section{Sample description}\label{sec:sample}
In order to estimate the volumetric densities, we need a sample of star-forming galaxies with known gravitational potential and their observed surface 
densities of gas and SFR as a function of galactocentric radius $R$. 
We selected the galaxies starting from the sample of The HI Nearby Galaxy Survey \citep[THINGS;][]{2008Walter}, which includes 34 objects.

\subsection{Surface densities}\label{sec:sample_surf}
Among the THINGS sample, we selected all the 23 galaxies in the sample of \citet{2008Leroy}, who provide the surface densities radial profiles for HI and SFR.  
\citet{2008Leroy} derived the atomic gas distribution from the THINGS 21 cm emission maps. 
The SFR distribution was obtained combining the far-ultraviolet (unobscured SF) emission maps from the Galaxy Evolution Explorer 
\citep[GALEX;][]{2007GildePaz} and the 24 $\mu$m (obscured SF) emission maps from the Spitzer Infrared Nearby Galaxy Survey \citep[SINGS;][]{2003Kennicutt}. 
These authors divided each galaxy in rings and calculated the surface densities at a certain radius as azimuthal averages inside that ring. 
This method is supposed to smooth the distributions and cancel azimuthal variations due to over- or under-dense regions as holes or spiral arms. 
\cite{2008Leroy} used the CO(2-1) transition maps from the HERA CO-Line Extragalactic Survey \citep[HERACLES;][]{2005Leroy} and the CO(1-0) transition 
maps from the Berkeley-Illinois-Maryland Association Survey Of Nearby Galaxies \citep[BIMA SONG;][]{2003Helfer} to calculate the H$_2$ 
surface densities for about half of the galaxies in their sample. 
These authors also used the MW $\alpha_\mathrm{CO}$ to convert the integrated CO intensity to H$_2$ surface density. 
However, as shown by \cite{2012Narayanan}, the choice of $\alpha_\mathrm{CO}$ is crucial as it influences the shape of SF laws, in particular at high surface 
density regimes. 
Hence, we took the profiles for molecular gas from \citet{2016Frank}, who used the same data as \cite{2008Leroy} but adopted 
the $\alpha_\mathrm{CO}$ factor reported by \citet{2013Sandstrom}. 
These authors took account of the dust-to-gas ratio and the metallicity gradient to obtain an accurate estimate of the $\alpha_\mathrm{CO}$ radial 
variation in 26 nearby galaxies. 
They found that the radial profile of $\alpha_\mathrm{CO}$ is nearly constant for all the galaxies, except in the central regions, where it tends 
to decrease and becomes 5-10 times smaller than the MW value in the most extreme cases. 
For example, the inner H$_2$ surface densities in NGC 4736 and NGC 5055 that were calculated by \cite{2013Sandstrom} differ from \cite{2008Leroy} 
results by one order of magnitude. 
For NGC 2403, \citet{2016Frank} used the MW $\alpha_\mathrm{CO}$ as this galaxy was not included in \cite{2013Sandstrom} study. 

\subsection{Selection based on mass models}\label{sec:sample_pot}
Among \cite{2008Leroy} sample, we selected the galaxies with parametric mass models in \cite{2008Deblok} or \citet{2016Frank}. 
In particular, \cite{2008Deblok} decomposed high quality HI rotation curves for a sample of 19 THINGS galaxies to obtain mass models using a DM halo, 
a stellar disc, a bulge (if present), and an atomic gas disc. 
Concerning the DM component, the authors adopted either an isothermal (Eq.~\ref{eq:isohalo}) or a NFW profile (Eq.~\ref{eq:nfwhalo}): in the 
first case they provide the best-fit central volume density $\rho_{\mathrm{DM},0}$ and core radius $r_\mathrm{c}$, while in the second case the 
parameters are the concentration $c$ and $V_{200}$, which is the circular velocity at $r_{200}$. 
For the stellar disc component, \cite{2008Deblok} fitted the 3.6 $\mu$m intensity profile with Eq.~\ref{eq:expsec_stardisc} leaving $R_\star$ and the 
mass-to-light ratio M/L as free parameters. 
In a small number of galaxies, they found an additional central component in the 3.6 $\mu$m surface brightness distribution, which is related to the 
stellar bulge. 
These authors fitted the light profile using the same profiles of the stellar disc (Eq.~\ref{eq:expsec_stardisc}) instead of a more generic Sersic profile $R^{1/n}$ 
\citep{1963Sersic}. 
The main reason for this choice was the limited radial range over which the bulge profile dominated the total emission and this avoided the need for the  
determination of the index $n$. 
They checked that assuming a different functional form did not significantly impact on their final mass models. 
Concerning the atomic gas component, \cite{2008Deblok} assumed that it is distributed in an infinitely thin disc. 
Later, \citet{2016Frank} repeated the \cite{2008Deblok} analysis including the molecular gas contribution for 12 galaxies; the molecular gas disc was also 
assumed to be infinitely thin in the modelling. 
Frank et al. found a good agreement with \cite{2008Deblok} results and improved the DM halo parametric mass model for some galaxies. 

Cross-matching \cite{2008Deblok} and \cite{2008Leroy} samples, we ended up with a sample of 12\footnote{We excluded NGC 3521 from our study as its HI disc 
shows a prominent warp along the line of sight. This feature complicates the analysis of the HI kinematics and the determination of its velocity dispersion.} 
nearby star-forming galaxies with surface densities of gas and SFR, and parametric mass models. 
In our sample, there are six normal spirals and six low-mass galaxies, whose circular velocity do not exceed 150 \kms (DDO 154 is a dwarf galaxy). 
DDO 154\footnote{In dwarf galaxies, the asymmetric-drift correction should be included in the determination of the rotation curve. 
However, \citet{2017Iorio} showed that its contribution is negligible in the case of DDO 154, which is the least massive galaxy in our sample.}, 
IC 2574, and NGC 7793 were not included in \citet{2016Frank} sample as no CO emission was detected, for these we used the mass models reported in 
\cite{2008Deblok}. 
The main properties of the galaxies and the parameters of their mass models are summarised in Tables~\ref{tab:props} and \ref{tab:gravpot_params_main}, respectively. 

For the sake of accuracy, we checked that the distances reported by \cite{2008Deblok} and \cite{2016Frank} are compatible with those 
reported in \cite{2016Lelli}, who carefully selected the most reliable measurements in literature (except for NGC 0925 and NGC 4736, which are not 
included in the \citealt{2016Lelli} sample). 
For some galaxies (DDO 154, IC 2574, NGC 5055, NGC 6946, and NGC 7793), the difference between the two distances is not negligible and could slightly 
influence the rotation curve. 
Hence, we decided to adopt \cite{2016Lelli} distances and correct the surface densities of \cite{2008Leroy} accordingly. 

\subsubsection{Galaxies with bulge}\label{sec:bulge}
As mentioned above, \cite{2008Deblok} modelled the mass distribution of the bulges using Eq.~\ref{eq:expsec_stardisc}, i.e. as they were exponential discs. 
This is not convenient for our purpose as the vertical pull near the midplane in the potential of this flattened component is stronger than the same 
force in the potential of a more realistic spheroidal distribution with the same mass.  
Therefore, the scale height would be significantly smaller, at least for the innermost regions where the bulge is likely the dominant component of the total 
gravitational potential. 
To alleviate this problem, we built the alternative bulge model described in Sec.~\ref{sec:bulge_model} using an exponential sphere 
(Eq.~\ref{eq:bulge_expball}). 
In this way, the observed exponential light distributions are preserved, but the mass distributions are no more flattened across the midplane. 

We want our model of the exponential sphere to have the same circular velocity as the (bulge) model of exponential disc of \cite{2008Deblok} 
for each galaxy with significant contribution from the bulge (NGC 2841, NGC 4736, NGC 5055, NGC 6946, and NGC 7331). 
To this purpose, we fitted the circular velocity of the exponential sphere
\begin{equation}\label{eq:vcirc_expball}
 V_\mathrm{c,b}(r)= \sqrt{ 4 \pi G \rho_\mathrm{b,0} \frac{r_\mathrm{b}}{r} 
 \left[ 2 r_\mathrm{b}^2 -  
 \left( r^2 + 2 r r_\mathrm{b} + 2 r_\mathrm{b}^2 
 \right) e^{-r/r_\mathrm{b}} \right] }
\end{equation}
to the circular velocity of \cite{2008Deblok} flat bulge, leaving $\rho_\mathrm{b,0}$ and $r_\mathrm{b}$ as free parameters. 
In the end, our models for the bulges are given by Eq.~\ref{eq:bulge_expball} with the best-fit $\rho_\mathrm{b,0}$ and $r_\mathrm{b}$ reported in 
Table~\ref{tab:gravpot_params_main}. 

\renewcommand{\arraystretch}{1.5}
\begin{table}
        \centering
        \caption{Properties of the sample galaxies: 
        (1) morphological type;
        (2) distance; 
        (3) mean value of the flat part of the rotation curve \citep[][except for NGC 0925, NGC 4736, and NGC 7793, see 
        Appendices~\ref{ap:3DB} and~\ref{ap:ngc7793} for details]{2016Lelli};
        (4) inclination; 
        (5) position angle.}
        \label{tab:props}
        \begin{tabular}{l l c c c c}
        \hline\hline
        Galaxy  & Type          & D     & V$_\mathrm{flat}$     & $i$                   & P.A.            \\
                &               & (Mpc) & (\kms)                & (\textdegree)                 & (\textdegree)           \\
                & (1)           & (2)   & (3)                   & (4)                   & (5)             \\
        \hline
        DDO 154 & Im            & 4.04  & 47.0                  & 65.0                  & 224.0           \\
        IC 2574 & Sm            & 3.91  & 66.4                  & 53.0                  & 56.0            \\
        NGC 0925& SABd          & 9.20  & 117.5                 & 58.0                  & 287.0           \\
        NGC 2403& Scd           & 3.16  & 131.2                 & 61.0                  & 124.5           \\
        NGC 2841& Sb            & 14.10 & 284.8                 & 73.7                  & 152.6           \\
        NGC 2976& Sc            & 3.58  & 85.4                  & 61.0                  & 334.5           \\
        NGC 3198& Sc            & 13.80 & 150.1                 & 71.5                  & 216.0           \\
        NGC 4736& SABa          & 4.70  & 151.7                 & 41.4                  & 306.7           \\
        NGC 5055& Sbc           & 9.90  & 179.0                 & 55.0                  & 101.8           \\
        NGC 6946& Scd           & 5.52  & 158.9                 & 33.0                  & 243.0           \\
        NGC 7331& Sb            & 14.70 & 239.0                 & 75.8                  & 167.7           \\
        NGC 7793& Sd            & 3.61  & 121.8                 & 47.0                  & 290.1           \\
        \hline
        \end{tabular}
\end{table}

\begin{table*}
        \centering
        \caption{Parametric mass models for DM and stellar components of the sample galaxies.
        DM halo profile: 
        (1) type, ISO=isothermal, NFW=Navarro-Frenk-White;
        (2) concentration or central density; and
        (3) V$_{200}$ or core radius. 
        Stellar disc:
        (4) central surface density $\Sigma_{\star,0}=\rho_{\star,0}/(2R_\star)$;
        (5) scale radius;
and         (6) scale height. 
        Bulge (as exponential sphere): 
        (7) central density and        (8) scale radius.}
        \label{tab:gravpot_params_main}
        \begin{tabular}{l|ccc|ccc|cc}
        \hline\hline
        Galaxy                  &\multicolumn{3}{c|}{Dark matter halo}                          & \multicolumn{3}{c|}{Stellar disc}               & \multicolumn{2}{c}{Bulge}     \\
                                & Type  & $c$ -- $\rho_\mathrm{DM,0}$   & V$_{200}$ -- $r_\mathrm{c}$     & $\Sigma_{\star,0}$    & $R_\star$     & $z_\star$       & $\rho_\mathrm{b,0}$   & $r_\mathrm{b}$\\
                                &       & (const -- $10^6$M$_\odot$kpc$^{-3}$)  & (\kms -- kpc)           & ($10^6$M$_\odot$kpc$^{-2}$)& (kpc)    & (kpc)         & ($10^{10}$M$_\odot$kpc$^{-3}$)& (kpc)   \\
                                & (1)   & (2)                           & (3)                             & (4)                   & (5)           & (6)             & (7)                   & (8)           \\
        \hline
        DDO 154                 & ISO   & 28.5                          & 1.32                            & 5.7                   & 0.72          & 0.144           & 0                     & 0             \\
        IC 2574                 & ISO   & 5.0                           & 6.18                            & 14.5                  & 2.85          & 0.57            & 0                     & 0             \\
        NGC 0925                & ISO   & 6.5                           & 8.90                            & 68.6                  & 4.1           & 0.82            & 0                     & 0             \\
        NGC 2403                & ISO   & 144.4                         & 1.50                            & 176.4                 & 1.81          & 0.362           & 0                     & 0             \\
        NGC 2841                & NFW   & 24.8                          & 172.6                           & 684.4                 & 4.2           & 0.84            & 1.24                  & 0.394         \\
        NGC 2976                & ISO   & 42.8                          & 2.60                            & 247.4                 & 0.9           & 0.18            & 0                     & 0             \\
        NGC 3198                & ISO   & 45.2                          & 2.80                            & 302.3                 & 3.06          & 0.612           & 0                     & 0             \\
        NGC 4736                & NFW   & 108.3                         & 40.9                            & 529.8                 & 1.99          & 0.398           & 5.3                   & 0.144         \\
        NGC 5055                & ISO   & 11.1                          & 7.15                            & 1179.0                & 3.2           & 0.64            & 3.0                   & 0.19          \\
        NGC 6946                & ISO   & 31.4                          & 4.80                            & 752.2                 & 2.97          & 0.594           & 21.0                  & 0.08          \\
        NGC 7331                & NFW   & 9.3                           & 171.2                           & 1160.9                & 3.3           & 0.66            & 10.8                  & 0.175         \\
        NGC 7793                & ISO   & 93.5                          & 1.95                            & 420.7                 & 1.3           & 0.26            & 0                     & 0             \\
        \hline
        \end{tabular}
\end{table*}
%
\section{Gas disc thickness}\label{sec:thick}
In this section, we calculate the scale height of HI, H$_2$, and SFR distributions. 
As mentioned in Section~\ref{sec:model}, the vertical distribution of the gas (Eq.~\ref{eq:densprof}) is regulated by the total gravitational potential 
of the galaxy and the gas velocity dispersion, which have opposite roles. 
The main obstacle to the scale height calculation is accounting for the gas self-gravity.
Indeed, the total gravitational potential of a galaxy $\Phi$ must include also the gas contribution, which depends on the gas distribution itself and thus 
on the scale height. 

In order to include the self-gravity, we used the publicly available software \textsc{Galpynamics}\footnote{\url{https://github.com/iogiul/galpynamics}} 
\citep{2018Iorio} to compute the gas potential and scale height through an iterative algorithm \citep[see also][]{2008Abramova,2011Banerjee}, which we explain in this section in a broad outline. 
In order to choose a simple example, let us consider a galaxy composed of DM, stars and atomic gas (including He). 
\begin{enumerate}
 \item As a preliminary stage, the software calculates the potential of DM and stars, which is defined as the external and fixed potential 
 $\Phi_\mathrm{ext}$. 

 \item In the zero-order step, \textsc{Galpynamics} assumes a razor-thin ($h_\mathrm{HI}=0$) mass distribution for the HI disc and calculates its 
 gravitational potential $\Phi_\mathrm{HI}$. The total gravitational potential of the galaxy is then set to $\Phi = \Phi_\mathrm{ext}+\Phi_\mathrm{HI}$. 
 
 \item The first iteration begins. The HI vertical profile is given by Eq.~\ref{eq:densprof}, where the velocity dispersion is given by 
 Eq.~\ref{eq:exp_vdisp}, and it is fitted with a Gaussian function (Eq.~\ref{eq:densprofscale})  to infer the new scale height $h_\mathrm{HI}^{'}$. 
 The next evaluation of the HI gravitational potential $\Phi^{'}_\mathrm{HI}$ is done for a disc with thickness $h_\mathrm{HI}^{'}$. 
 Then, we are able to update the total potential to $\Phi^{'} = \Phi_\mathrm{ext}+\Phi_\mathrm{HI}^{'}$. 
 
 \item Using $\Phi^{'}$ in Eq.~\ref{eq:densprof}, we find more accurate vertical distribution and scale height $h_\mathrm{HI}^{''}$ for the atomic gas, 
 which allow us to better estimate  $\Phi^{''}_\mathrm{HI}$ and then $\Phi^{''}$. 
\end{enumerate}
This procedure is iterated until two successive computations of the scale height differ by less than a tolerance factor, chosen by the user. 
This software was extensively tested using mock data \citep[see][]{2018Iorio}.  

Most of the galaxies in our sample have both the atomic and molecular gas components. 
We first calculate the HI scale height in the gravitational potential of stars and DM, and then the scale height for H$_2$ but 
including also the HI gravitational potential. 
This choice implies that the HI distribution is not influenced by the H$_2$ distribution and that we obtain two different scale heights for each gas phase, 
$h_\mathrm{HI}$ and $h_{\mathrm{H_2}}$. 
We expect that including the molecular gas distribution to the potential does not affect the HI scale height, as the total mass of molecular gas is about 
one order of magnitude smaller than the total amount of atomic gas \citep[see][]{2008Leroy}. 
Moreover, the molecular phase is concentrated in the inner regions of galaxies, where stars are the dominant mass component, and becomes negligible in the 
outskirts. 
On the other hand, the atomic gas is distributed out to larger radii, so its contribution to the total gravitational potential there could become truly 
significant. 

\subsection{Flaring HI disc}\label{sec:hidisc}
In order to calculate the HI scale height, \textsc{Galpynamics} needs, in addition to the external potential $\Phi_\mathrm{ext}$, the HI radial profiles of 
the surface density $\Sigma_\mathrm{HI}(R)$ and velocity dispersion $\sigma_\mathrm{HI}(R)$.

\subsubsection{HI surface density}\label{sec:hidens}
\begin{figure*}
\includegraphics[width=2.\columnwidth]{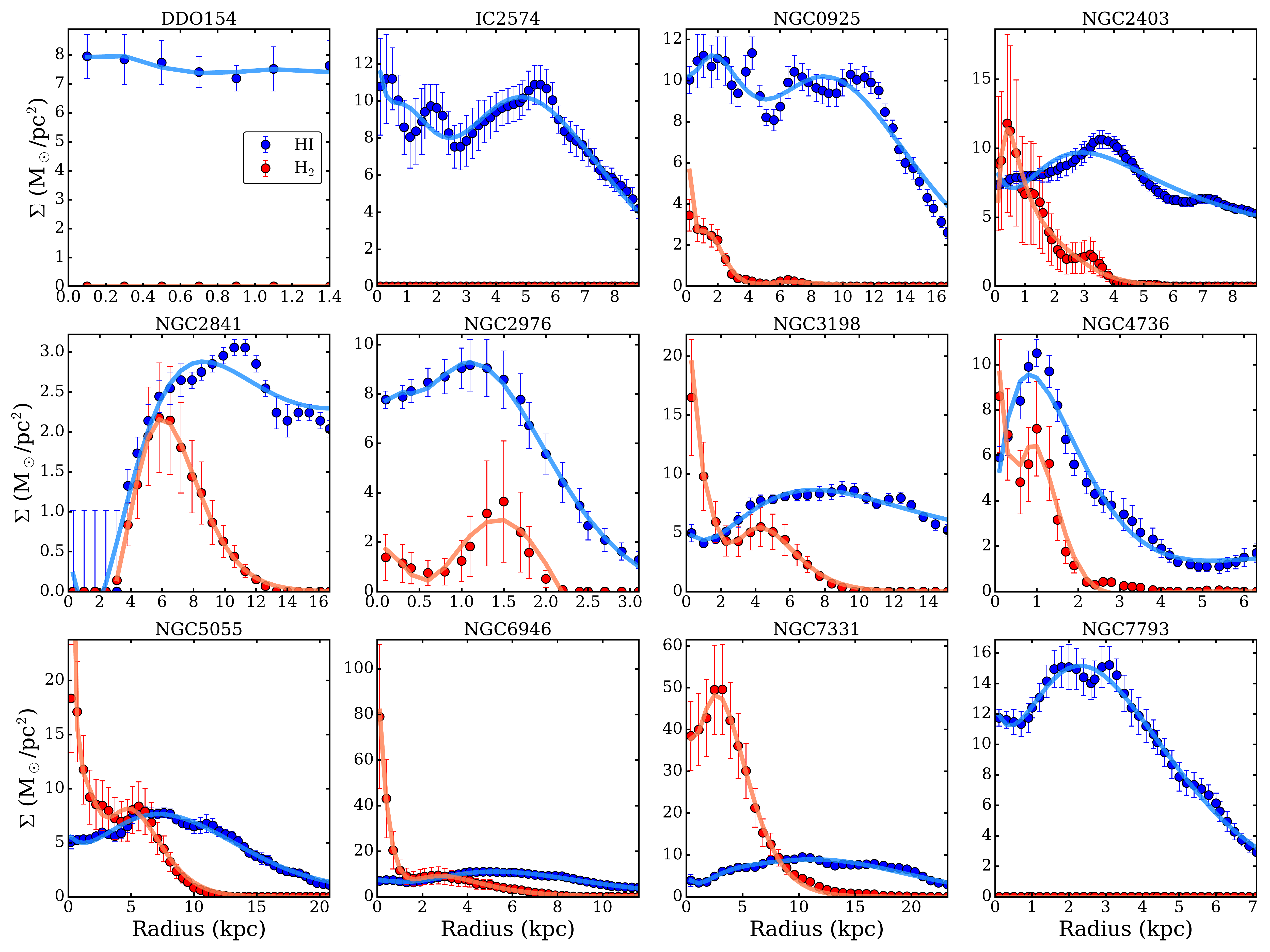}
\caption{HI  \citep[blue points]{2008Leroy} and H$_2$ \citep[red points]{2016Frank} observed surface densities radial profiles. 
The light blue and coral curves show the models used to compute the scale heights and are obtained by fitting  Eq.~\ref{eq:poly} to the observed profiles. 
Only the ranges where the SFR is measured are shown.}
\label{fig:plottone_allsurfdens_fits}
\end{figure*}

As mentioned in Sec.~\ref{sec:gas_model}, we modelled the atomic gas distribution using a combination of an exponential and a polynomial (Eq.~\ref{eq:poly}), 
which was fitted on the observed azimuthally averaged radial profiles of \cite{2008Leroy} leaving $\Sigma_\mathrm{HI,0}$, $R_\Sigma$, and $C_i$ as free 
parameters (the helium correction of 1.36 is included). 
In Fig.~\ref{fig:plottone_allsurfdens_fits}, the observed $\Sigma_\mathrm{HI}(R)$ for each galaxy is shown by the blue points and the corresponding best-fit 
model is represented by the light blue curve. 
It is clear that the best fits reproduce well the observed radial profiles save negligible and small differences, which do not affect the computation of 
the scale height. 

\subsubsection{HI velocity dispersion}\label{sec:hivdisp}
As shown by Eq.~\ref{eq:densprof}, we expect the scale height to linearly depend on the velocity dispersion, so an accurate modelling of 
$\sigma_\mathrm{HI}(R)$ radial profile is fundamental. 
To this aim, we derived the radial profiles of the velocity dispersion in our galaxies using the publicly available software $^{\mathrm{3D}}$\textsc{Barolo}
\footnote{\url{http://editeodoro.github.io/Bbarolo/}} \citep{2015Diteodoro}, hereafter 3DB, on THINGS data cubes \citep{2008Walter}. 
The 3DB software performs a tilted-ring model fitting directly on the data cube, allowing us to correct for the beam smearing, which can significantly modify the 
resulting velocity dispersion and rotation curve \citep[e.g.][]{1999PhDTSwaters}. 
Moreover, the rotation velocity and velocity dispersion are fitted simultaneously rather than as separate components, as done in the classical 2D 
approach based on velocity dispersion maps \citep[e.g.][]{2009Tamburro,2017Romeo}. 
We chose 400 pc as a common spatial resolution for the data cubes of our galaxies, which is a compromise between negligible gas streaming motions within our beam 
and sufficient signal-to-noise ratio (S/N) in low column density areas. 
Details on the properties of the data cubes and the 3DB input parameters are found in Appendix~\ref{ap:3DB}. 
Fig.~\ref{fig:vdisp_fits} shows the velocity dispersion measured by 3DB for all the galaxies in the sample. 
Our results are in agreement with previous works showing that the velocity dispersion decreases with the radius from 12-20 \kms in the inner parts of 
local spirals and dwarfs down to 5-7 \kms in the outskirts \citep[e.g.][]{2002NarayanJog,2008Boomsma,2009Tamburro}. 
\begin{figure*}
\includegraphics[width=2.\columnwidth]{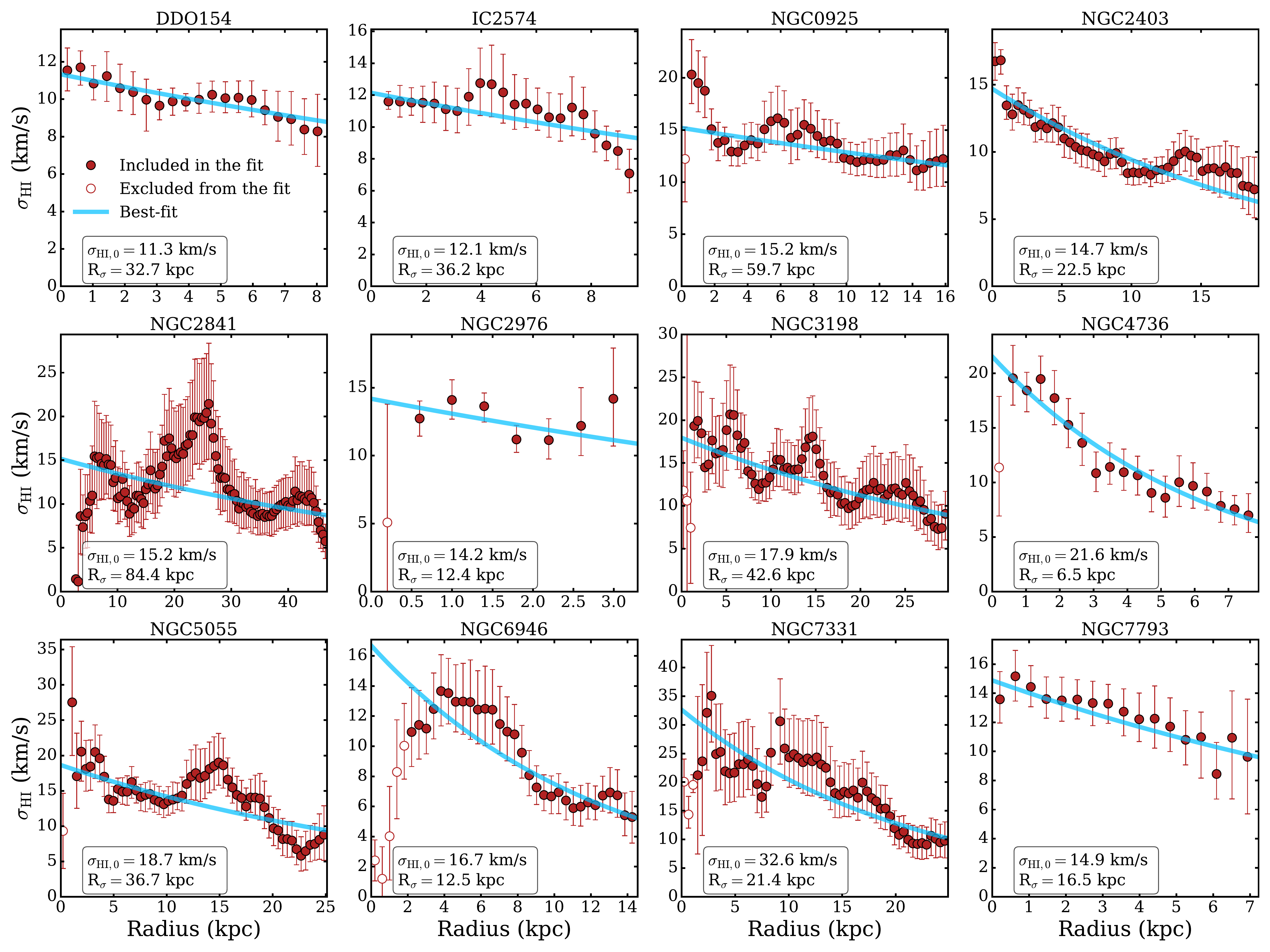}
\caption{HI velocity dispersion measured using $^{\mathrm{3D}}$\textsc{Barolo} (filled and empty circles) for our galaxies with a common sampling 
of about 400 pc (for NGC 2841 we show one data point every two). 
The best-fit models (Eq.~\ref{eq:exp_vdisp}) are shown by the light blue curves, whose parameters ($\sigma_{\mathrm{HI},0}$ and $R_\sigma$) are reported in 
a box in the lower left corner of each panel. The points indicated by empty circles are excluded from the fit.}
\label{fig:vdisp_fits}
\end{figure*}

The velocity dispersion of NGC 7331 galaxy is probably overestimated. 
Indeed, if we compare this profile to that of the other galaxies, we see that it is systematically higher. 
This increase likely originates from projection effects due to the galaxy inclination angle and the HI disc thickness or non-circular motions along the 
line of sight, which bias the velocity dispersion towards high values. 
NGC 7331 is indeed the most inclined galaxy in our sample ($i\approx$ 76\textdegree), so the line of sight intercepts regions with different rotation 
velocity, broadening the line profile. 
Such effects may be present in two further profiles of very inclined galaxies, NGC 2841 and NGC 3198, but they seem to be less affected. 
NGC 2841 velocity dispersion shows a peculiar sharp increase of 10 \kms extending from 15 kpc to 30 kpc, whose origin we discuss in 
Appendix~\ref{ap:ngc2841_xshape}.

Having measured $\sigma_\mathrm{HI}(R)$, the model for the velocity dispersion (Eq.~\ref{eq:exp_vdisp}) was fitted to the data points leaving 
$\sigma_\mathrm{HI,0}$ and $R_\sigma$ as free parameters. 
The model must reproduce the radial decrease of the velocity dispersion, leaving aside the most peculiar features differing from the global trend, which 
could be due to low S/N regions or some residual beam smearing effect in the very innermost radii. 
Therefore, we excluded the innermost point of NGC 0925, NGC 2841, NGC 2976, NGC 3198, NGC 4736, NGC 5055, and NGC 7331. 
For NGC 6946, we rejected the inner five velocity dispersion measurements after a comparison with the velocity dispersion profile of 
\cite{2008Boomsma}, who found that $\sigma_{\mathrm{HI}} \approx 12-15$ \kms for the central radii. 
Thus, the drop that we observe is likely an artefact due to low S/N of our data, which have higher angular resolution with respect to \citeauthor{2008Boomsma}. 
In Fig.~\ref{fig:vdisp_fits}, the excluded points are shown as empty circles, while the measurements used for the fit are shown as the filled circles.

\subsubsection{HI scale height}\label{sec:hiscaleheight}
\begin{figure}
\includegraphics[width=1.\columnwidth]{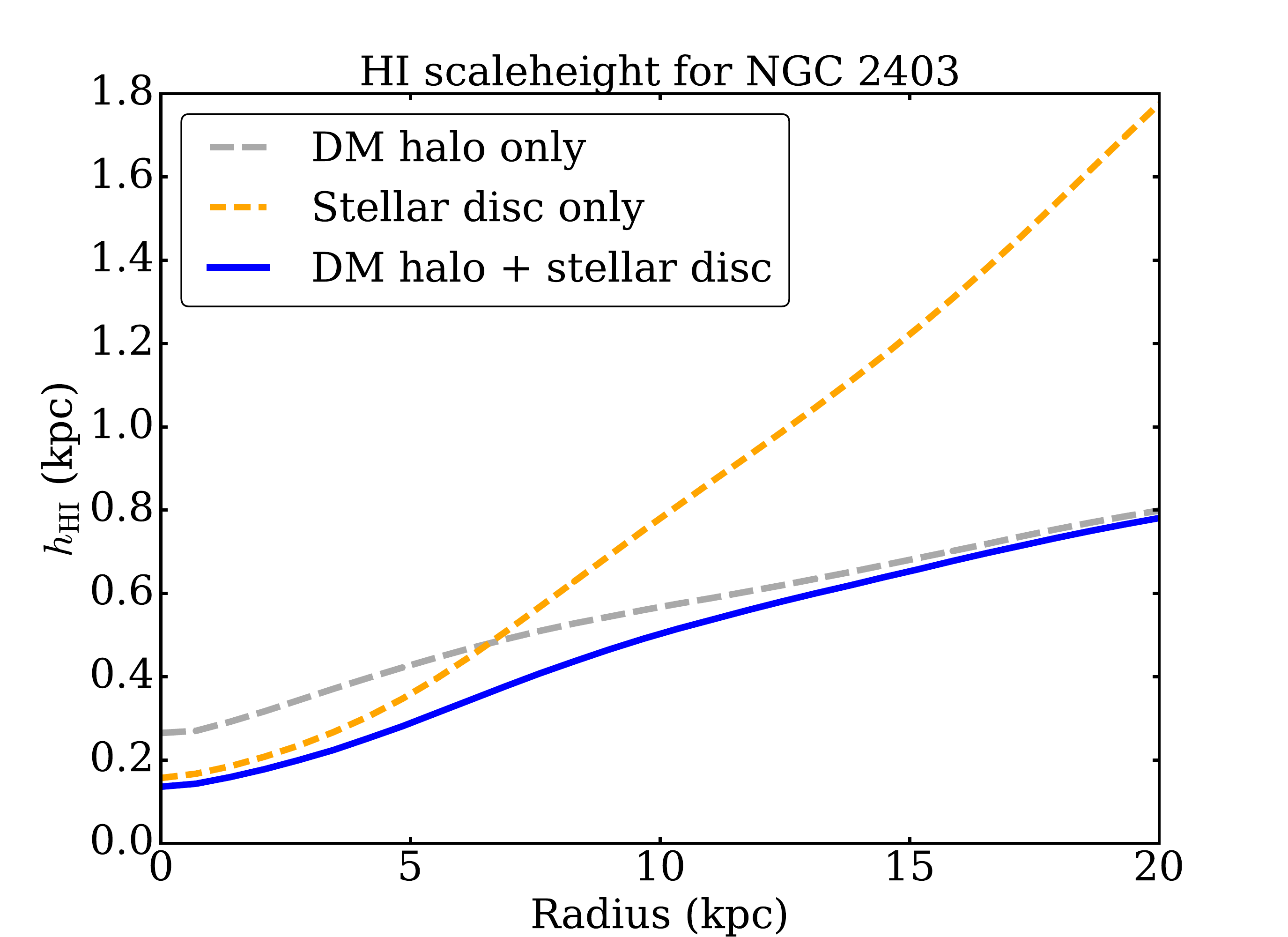}
\caption{HI scale height radial profiles in the presence of three different gravitational potentials (including the HI self-gravity) but with the same 
velocity dispersion. 
The orange and grey dashed curves, respectively, show $h_\mathrm{HI}(R)$ for the stellar disc only and DM halo only potentials 
taken from NGC 2403 mass model. 
The total potential of NGC 2403 gives the blue solid curve.}
\label{fig:h_sepcomp}
\end{figure}

We calculated $h_{\mathrm{HI}}(R)$ for our galaxies using their gravitational potential and the surface density and velocity dispersion of the atomic gas. 
Before describing the full sample, it is useful to focus on a single galaxy in order to understand which mass component drives the trend of the 
scale height with radius. 
In Fig.~\ref{fig:h_sepcomp}, we show three different HI scale heights out to $R=20$ kpc for NGC 2403: each of these scale heights is obtained with a different 
gravitational potential but the same velocity dispersion radial profile $\sigma_\mathrm{HI}(R)$.
In the presence of the stellar disc only (dashed orange line), the scale height increases exponentially out to about $R=7$ kpc, then the growth 
becomes milder and $h_\mathrm{HI}$ reaches 1.8 kpc at $R=20$ kpc. 
This is because the disc mass distribution fades within a short length, so the gravitational pull towards the midplane quickly weakens. 
As a consequence, the HI disc becomes thicker and thicker with radius, despite the decrease of the velocity dispersion; if the velocity 
dispersion were constant, then the flaring would be more prominent. 
For the DM only potential (dashed grey line), the pull towards the midplane is still significant in the outskirts, as the radial decrease of the DM density 
is significantly slower with respect to an exponential profile. 
In the combined potential of stars and DM (solid blue), the scale height is mainly driven by the stellar disc in the inner regions and by the 
DM halo in the outskirts \citep[see also][]{2018Sarkar}. 
At the end, $h_\mathrm{HI}$ increases by a factor of about 8 within 20 kpc in radius. 
We note that the scale  height in the single component potentials is always larger with respect to the combined potential, so 
neglecting one or the other component causes an overestimate of the scale height. 

The radial profiles of the HI scale height for the all galaxies in our sample are shown by the blue curves in Fig.~\ref{fig:plot_scaleheights}, and the associated 
uncertainties are represented by the faded blue area. 
In Appendix~\ref{ap:error_prop}, we provide details about the estimates of the uncertainties, which include the errors on $\Sigma_\mathrm{HI}$ and 
$\sigma_\mathrm{HI}$. 
We note the global trend of the flaring is similar for all the galaxies. 
We emphasise that the HI disc flaring is significant, regardless of the galaxy type, so assuming a thin gaseous disc or a constant thickness is never a 
good approximation. 
The presence of the bulge (NGC 2841, NGC 4736, NGC 5055, NGC 6946, and NGC 7331) reduces the scale height in the innermost regions. 
However, the mass model for the bulge is more uncertain (see Sec.~\ref{sec:bulge_model}) and the velocity dispersion in the centre of galaxies has large 
errors, so it is likely that the scale height in the innermost radii of these galaxies is underestimated or at least uncertain. 
The projection effects are particularly significant in the outskirts, therefore we expect that the intrinsic volume densities distribution with radius will differ
from the observed surface densities distribution. 
Therefore, we anticipate that the VSF law will have different shape than the law based on surface densities. 
\begin{figure*}
    \includegraphics[width=2\columnwidth]{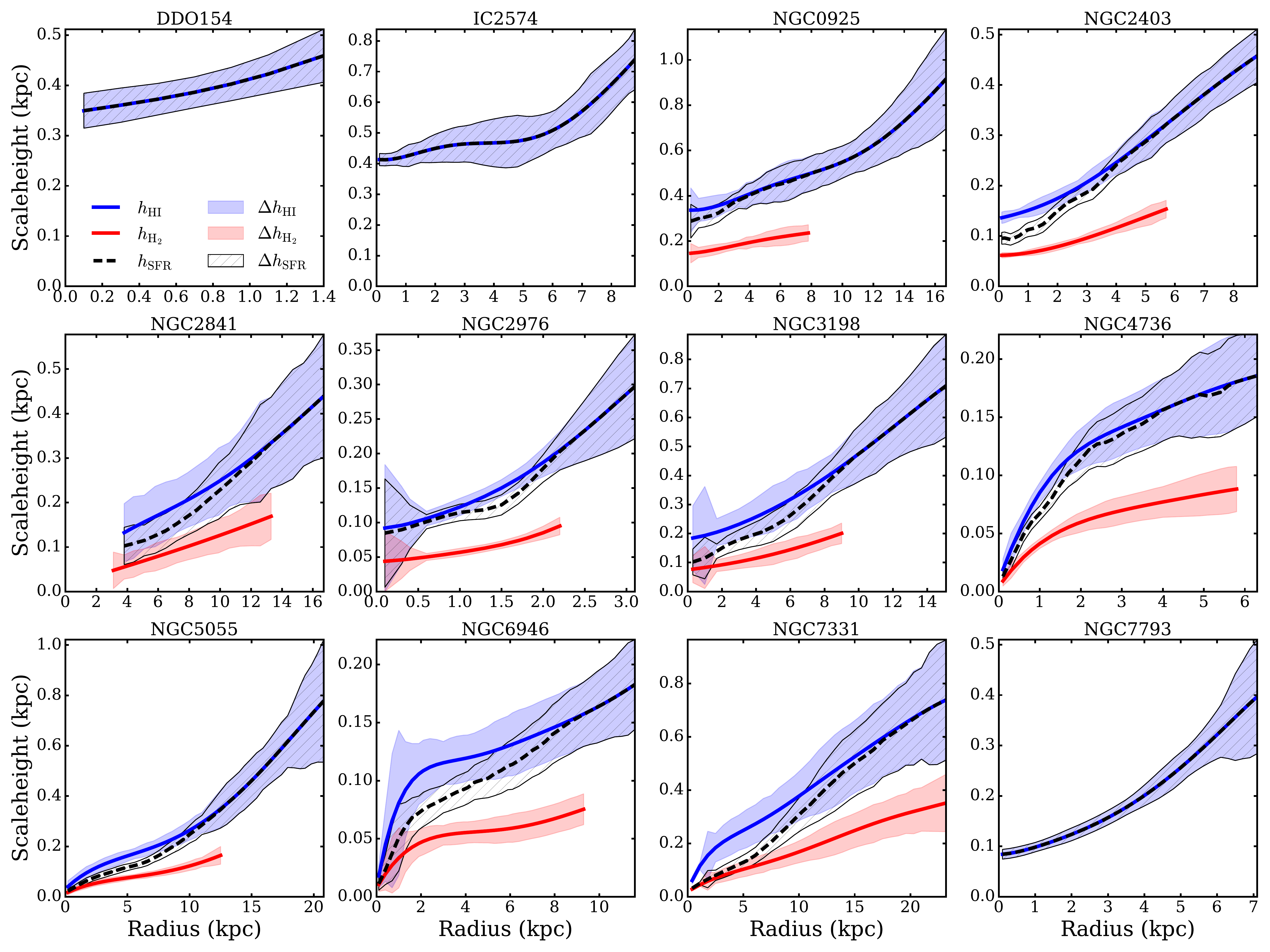}
    \caption{Scale height radial profiles: $h_\mathrm{HI}(R)$ (solid blue) and $h_\mathrm{H_2}(R)$ (solid red) are calculated by \textsc{Galpynamics}, while 
    $h_\mathrm{SFR}(R)$ (dashed black) is estimated using Eq.~\ref{eq:hSFR_wmean}. 
    We note that $h_\mathrm{H_2}(R)$ is shown out to the radius where CO emission is detected. 
    The faded regions indicate the uncertainties on the gas scale heights, while the black dashed regions are the uncertainties on $h_\mathrm{SFR}(R)$.}
    \label{fig:plot_scaleheights}
\end{figure*}

\subsection{Flaring H$_2$ disc}\label{sec:h2disc}
The molecular gas scale height was estimated using the gravitational potential of stars, DM, and the HI disc with flaring thickness. 
Moreover, we needed the surface density and velocity dispersion of the molecular gas. 

\subsubsection{H$_2$ surface density}\label{sec:h2dens}
As in Sec.~\ref{sec:hidens}, the model for the H$_2$ distribution (Eq.~\ref{eq:poly}) was fitted to the radial profile of the observed surface density 
(including the correction for Helium), leaving $\Sigma_\mathrm{H_2,0}$, $R_\Sigma$, and $C_i$ as free parameters. 
In Fig.~\ref{fig:plottone_allsurfdens_fits}, the observed $\Sigma_\mathrm{H_2}$ from \cite{2016Frank} are shown by the red points and the corresponding 
best-fit models are represented by the coral curves. 
The error bars include the uncertainties on $\alpha_\mathrm{CO}$ as reported in \cite{2013Sandstrom}. 

\subsubsection{H$_2$ velocity dispersion}\label{sec:h2vdisp}
As for $\sigma_\mathrm{HI}$ (Sec.~\ref{sec:hivdisp}), we modelled $\sigma_\mathrm{H_2}$ using an exponential profile, which should be fitted 
to the observed velocity dispersion radial profiles. 
The spatial and  spectral resolution are crucial in measuring the molecular gas velocity dispersion, as they could artificially broaden the 
observed emission line. 
The first typically affects the velocity dispersion in the central parts of the galaxies and the second acts as an additional broadening component. 

There are scarce high resolution observations of the molecular gas emission in our sample galaxies. 
In the literature, three studies measured the CO velocity dispersion radial profile using the HERACLES data cubes of the CO(2-1) emission 
line with spatial resolution of 13\arcsec. 
In particular, \cite{2013Caldu} used the data cubes with spectral resolution of 2.6 \kms and stacked the CO (and also HI) line profiles over kiloparsec-sized 
regions to improve the S/N. 
They used the HI velocity fields as a guide to align the profile centroids and measured the velocity dispersions by fitting a Gaussian function 
to the stacked profiles. 
They estimated the ratio of the HI to H$_2$ velocity dispersion to be  $\sigma_\mathrm{HI} / \sigma_{\mathrm{H}_2} \approx 1$. 
However, the staking method easily introduces an artificial broadening if the profiles are not perfectly aligned, so their result could be overestimated. 
Later, \cite{2016Mogotsi} \citep[see also][]{2017Romeo} used Hanning smoothed data cubes with spectral resolution of 5.2 \kms and fitted a Gaussian function 
to the line profiles in each pixel with S/N$>4$, finding $\sigma_\mathrm{HI} / \sigma_{\mathrm{H}_2} \approx 1.4$. 
Unfortunately, their resolution ($\approx 5$ \kms) is probably too low to measure the molecular gas velocity dispersion in the galaxy outskirts, where it can easily drop 
below 5 \kms as shown by \cite{2017Marasco}. 
These latter authors used the Leiden-Argentine-Bonn (LAB) all-sky 21 cm survey \citep{2005Kalberla} and the CO(2-1) survey 
\citep{2001Dame} to measure the distribution and kinematics of atomic and molecular gas with spectral resolution of about 2 \kms.  
They reproduced the observed emission building a model of the Galactic disc made of concentric and co-planar rings defined by rotation velocity, 
velocity dispersion, midplane volume density, and scale height. 
\cite{2017Marasco} showed that the radial trends of $\sigma_\mathrm{HI}$ and $\sigma_{\mathrm{H}_2}$ are approximetely the same \citep[see also][]{2016Mogotsi}, while 
their mean values are $8.9 \pm 1.5 $ \kms and $4.4 \pm 1.5 $ \kms, respectively. 

Hence, we decided to assume $\sigma_\mathrm{HI} / \sigma_{\mathrm{H}_2} \approx 2$ and estimated the radial profile of $\sigma_\mathrm{H_2}$ 
from the $\sigma_\mathrm{HI}$ radial profiles. 
In practice, the model for the molecular gas velocity dispersion is given by Eq.~\ref{eq:exp_vdisp} with $\sigma_{\mathrm{H}_2,0} =0.5 \sigma_\mathrm{HI,0}$ 
and the same $R_\sigma$ reported in Fig.~\ref{fig:vdisp_fits}. 
However, we tested that assuming 1.4 for $\sigma_\mathrm{HI} / \sigma_{\mathrm{H}_2}$ does not significantly affect our results. 
For completeness, we also compared our $\sigma_{\mathrm{H}_2}$ radial profile for NGC 2403, NGC 4736, and NGC 5055 with those reported by \cite{2011Wilson}. 
They measured $\sigma_{\mathrm{H}_2}$ using CO(3-2) emission data cubes with spectral resolution of 0.43 \kms and spatial resolution of 14.5\arcsec. 
Our profiles are compatible within the uncertainties with \citeauthor{2011Wilson} results save for the very central regions ($< 1-2$ kpc), 
where the beam smearing likely acts as an additional broadening component on their profiles. 

\subsubsection{H$_2$ scale height}\label{sec:h2scaleheight}
In Fig.~\ref{fig:plot_scaleheights}, we show the H$_2$ scale heights with their associated uncertainty for all the galaxies in our sample. 
In Appendix~\ref{ap:error_prop}, we explain how the errors on $h_{\mathrm{H}_2}$ were estimated to take account of the uncertainties on 
$\sigma_{\mathrm{H}_2}$ and $\Sigma_{\mathrm{H}_2}$, which include the error on $\alpha_\mathrm{CO}$. 
We note that $h_{\mathrm{H}_2} \approx 0.5 h_\mathrm{HI}$, save for negligible discrepancies, as the main driver of the difference in the flaring of HI and H$_2$ discs
is the velocity dispersion. 

\subsection{Star formation rate scale height}\label{sec:sfrscaleheight}
Knowing the scale heights of the HI and H$_2$, we estimated the scale height of the SFR vertical distribution using Eq.~\ref{eq:hSFR_wmean}. 
In Fig.~\ref{fig:plot_scaleheights}, we show $h_\mathrm{SFR}(R)$ (black dashed curve) as a function of radius and its uncertainties 
(see Appendix~\ref{ap:error_prop} for details). 
Clearly, in the case of DDO 154, IC 2574, and NGC 7793, $h_\mathrm{SFR}(R)$ coincides with $h_\mathrm{HI}$ as CO emission is not detected. 

%
\section{Volumetric star formation laws}\label{sec:vsf}
Having all the scale heights, we converted surface densities to volume densities through Eq.~\ref{eq:midplane}. 
In Appendix~\ref{ap:error_prop}, we describe the calculation of the uncertainties on the volume densities, which include the errors on the observed surface 
densities and on the scale heights. 

Fig.~\ref{fig:NGC5055colorsgals} illustrates the effect of the conversion to volume densities on the correlation between gas and SFR for the galaxy NGC\,5055. 
The left panel shows the classical surface-based correlation with each point coloured according to the radius. 
As can be seen from the central panel, the conversion of gas surface densities to volume densities using the constant $h_\mathrm{SFR}$ stretches 
the points along the $x$-axis.  
Indeed, low density points typically belong to the outskirts, therefore they undergo the most significant leftward shift. 
In this case, the SFR surface density profile is divided by a constant value, so its trend is not modified. 
In the right panel, the gas volume densities are the same as in the central panel, but the flaring $h_\mathrm{SFR}(R)$ (Fig.~\ref{fig:plot_scaleheights}) 
is assumed, so the points are also stretched along the $y$-axis. 

\subsection{Relation between total gas and star formation rate}\label{sec:totgas_corr}
\begin{figure*}
\centering
\subfloat[][\label{fig:NGC5055colorsgals}]
{\includegraphics[width=1\textwidth]{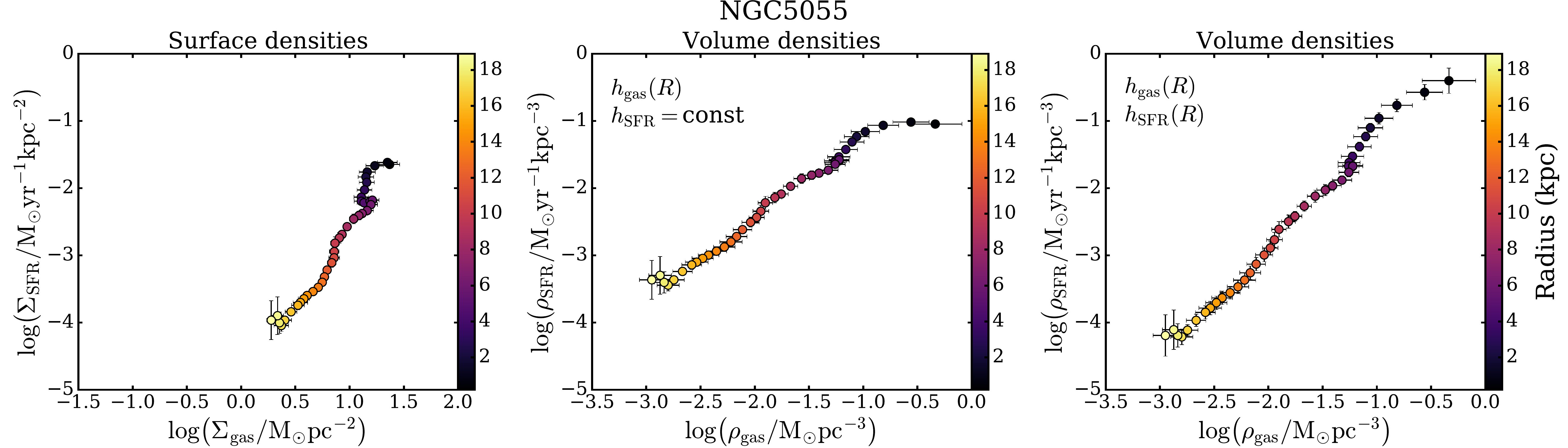}}
\\
\subfloat[][\label{fig:colorsgals}]
{\includegraphics[width=1\textwidth]{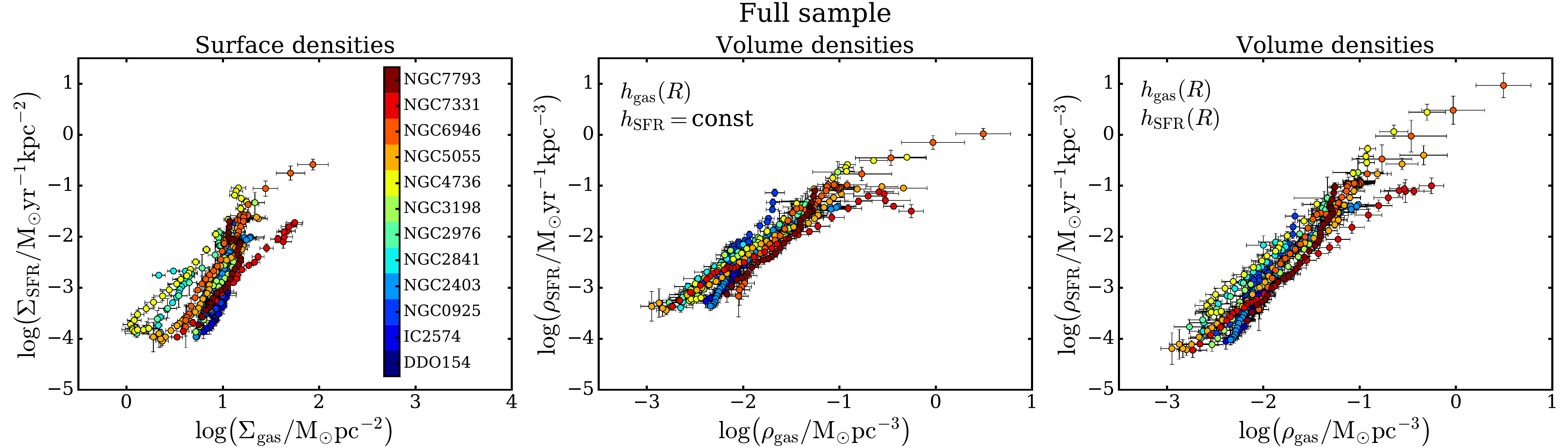}}
\caption{\textit{Upper row}: Correlations between gas and SFR surface densities (left) and volume (centre and right) densities for NGC 5055; 
    $h_\mathrm{SFR}$ is assumed to be constant and flaring (Eq.~\ref{eq:hSFR_wmean}) in the central and right panels, respectively. 
    Each point is obtained as an azimuthal average and coloured according to its galactocentric radius. 
    The slope of the VSF law is much shallower than for the surface-based law. This is a consequence of taking the flaring of the gas (and the SFR) 
    into account.
    \textit{Lower row}: Same as the upper row but for all the galaxies in our sample. Each galaxy has its own colour as shown by the colour bar. 
    The VSF law has considerably less scatter than the surface based version. 
    Each panel shows similiar ranges in $x$ and $y$. 
    No obvious break in the SF efficiencies is found at low densities after correcting for disc thickness.}
\label{fig:colorgals_names}
\end{figure*}

We now consider the full sample of galaxies. 
Fig.~\ref{fig:colorsgals} compares the surface-based (left) and the volume-based (centre and right) correlations between gas and SFR with the points colour-coded 
according to the galaxy of origin. 
By-eye, it is clear that the surface-based correlation is more scattered than any of the volume-based correlations. 
The change in the SF efficiency seen by \cite{2008Leroy} and \cite{2010Bigiel} is partially reduced in the left panel thanks to the improvement in 
the $\alpha_\mathrm{CO}$ measurement by \cite{2013Sandstrom} included in this study. 
However, some galaxies in the left panel (e.g. NGC 5055 and NGC 7793) seem to follow a steeper SF law with respect to the others (e.g. NGC 4736 and NGC 7331). 
Indeed, the observed surface density corresponds to the integral of the column of gas along the line of sight and the height of this gas column 
increases with radius.  
Hence, high surface densities can be present not only in the central parts of galaxies, but also in the external regions, where the volume density is instead low and 
a few stars form. 
On the other hand, using the volume densities, we found a tight correlation between SFR and gas over 4-5 orders of magnitude. 
Even by eye, it is clear that the scatter reduces as the galaxies tend to align on the top of each other. 

\begin{figure*}
\centering
\subfloat[][\label{fig:colorsgals_fgasrho}]
{\includegraphics[width=1\textwidth]{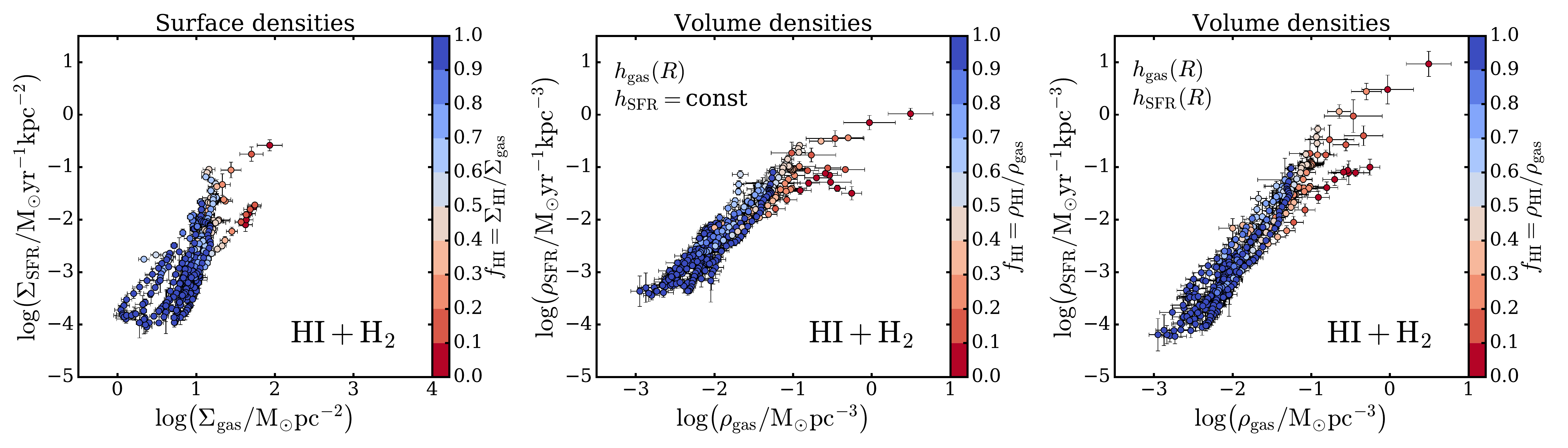}}
\\
\subfloat[][\label{fig:hist_gas}]
{\includegraphics[width=1\textwidth]{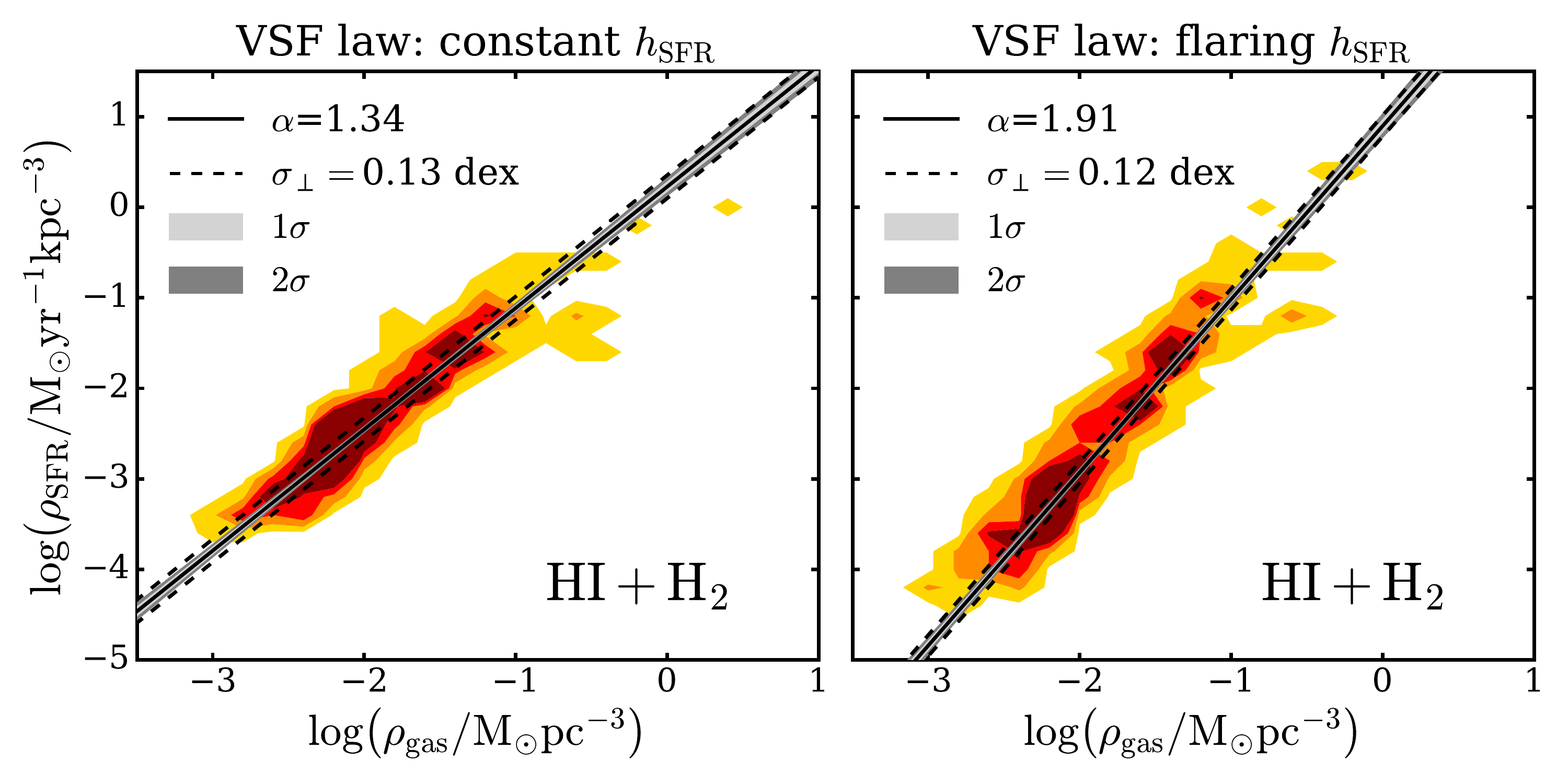}}
\caption{\textit{Upper row:} Same as Fig.~\ref{fig:colorsgals} but the points are colour-coded according to the HI fraction; blue and red points come from HI 
and H$_2$ dominated regions, respectively. The VSF law appears tight and straight even if only HI-dominated regions are considered. 
\textit{Lower row:} VSF law between total gas and SFR. The solid black line is the best-fit relation with slope $\alpha$ and orthogonal intrinsic scatter 
    $\sigma_\perp$ (dashed lines). The grey bands show $1\sigma$ and $2\sigma$ uncertainties on the fit.    
    In the left panel, $\rho_\mathrm{SFR}$ is calculated with the constant $h_\mathrm{SFR}$, while in the right panel $h_\mathrm{SFR}(R)$ 
    flares with radius (Eq.~\ref{eq:hSFR_wmean}). The volume densities radial profiles are shown as contours containing 95\% (yellow), 75\% (orange), 
    50\% (red), and 25\% (dark red) of the data points.}
\label{fig:fgas_cont_gas}
\end{figure*}

Fig.~\ref{fig:colorsgals_fgasrho} is the same as Fig.~\ref{fig:colorsgals} but the points are coloured according to the HI fraction, 
$f_\mathrm{HI}(R)=\Sigma_\mathrm{HI}(R) / \Sigma_\mathrm{gas}(R)$; the blue and red points are HI-dominated and H$_2$-dominated, respectively. 
Going from left to right along the $x$-axis of all panels, the molecular phase becomes more and more important, but the low density gas is mainly atomic. 
We note how the scatter in the HI-dominated regime is much reduced by the conversion from surface to volume densities. 

We then looked for a correlation between gas and SFR volume densities in the form of a power law as follows: 
\begin{equation}\label{eq:vsf_law}
 \rho_\mathrm{SFR} = A \rho_\mathrm{gas}^{\alpha} \, .
\end{equation}
The relation is univocally described by the normalisation $A$ and the index $\alpha$. 
We sampled the parameters space through the Monte Carlo - Marchov Chain (MCMC) method implemented in the Python package \texttt{emcee} 
\citep{2013ForemanMackey}. 
In logarithmic scale, the model is a simple linear relation with slope $\alpha$ and $y$-intercept $\log A$,
\begin{equation}\label{eq:vsf_law_log}
\log \rho_\mathrm{SFR} = \log A + \alpha \log \rho_\mathrm{gas} \, .
\end{equation}
We also included an intrinsic scatter, $\sigma_\perp$, which is orthogonal to the linear relation. 
We left slope, $y$-intercept, and scatter as free parameters in the Bayesian fit (see Appendix~\ref{ap:posteriors} for details). 
The case with constant $h_\mathrm{SFR}$ and that with flaring $h_\mathrm{SFR}(R)$ were studied separately. 
The best-fit parameters are reported in Table~\ref{tab:mcmc_results}; we found a slope of about 1.3 with $h_\mathrm{SFR}=100$ pc and about 
1.9 with the flaring $h_\mathrm{SFR}(R)$. 
This means that the slope of the VSF law cannot be univocally determined. 
However, if the true SFR scale height is between the two extreme choices, it is reasonable to think that also the true slope is between 1.3 and 1.9. 
The best-fit intrinsic scatter is very small in both cases ($\sigma_\perp \approx$ 0.1 dex). 
In Fig.~\ref{fig:hist_gas}, volume densities appear as contours and the panels show $\rho_\mathrm{SFR}$ in the constant (left) and the flaring 
$h_\mathrm{SFR}(R)$ (right) case. 
The best-fit relation is represented by the solid black line with the dashed lines showing $\pm \sigma_\perp$. 
In order to test the robustness of our results, we tried alternative formulations for $h_\mathrm{SFR}(R)$ as a function of the gas scale heights 
(e.g. harmonic mean) but the best-fit relations were compatible with those reported in Table~\ref{tab:mcmc_results} within the scatter of the VSF law. 

The high volume density regime above 0.1 M$_\odot$pc$^{-3}$ is the less sampled part of the plot and the scatter seems to increase there. 
Indeed, these points come from the innermost and H$_2$-dominanted regions of massive galaxies, where the $\alpha_\mathrm{CO}$ factor probably acts as an 
additional source of uncertainty on the surface density measurement. 
In particular, \cite{2013Sandstrom} discussed the reliability of their estimate of the $\alpha_\mathrm{CO}$ in the inner regions of galaxies, 
as they found that it is lower than the MW value and also well below the galaxy average. 
Out of a total of about 400 volume densities for our 12 galaxies, the H$_2$ fraction of only 25 points at most may be underestimated, so it is unlikely that 
our results would be influenced. 
As further test of the effect of $\alpha_\mathrm{CO}$ on the best fit, we repeated the whole procedure, including the scale heights calculation, using the H$_2$ 
surface densities of \cite{2008Leroy}, which were obtained assuming the MW $\alpha_\mathrm{CO}$ for all the galaxies. 
We found $\alpha=1.03 \pm 0.03$ and $\sigma_\perp=0.21 \pm 0.01$  in the case with the constant $h_\mathrm{SFR}$ and $\alpha=1.56 \pm 0.03$ and 
$\sigma_\perp=0.28 \pm 0.01$ with $h_\mathrm{SFR}(R)$, which is compatible with the relation in Fig.~\ref{fig:hist_gas}. 

\subsection{Atomic gas versus star formation rate}\label{sec:hionly_corr}
\begin{figure*}
\centering
\subfloat[][\label{fig:colorsgals_fgasrho_hi}]
{\includegraphics[width=1\textwidth]{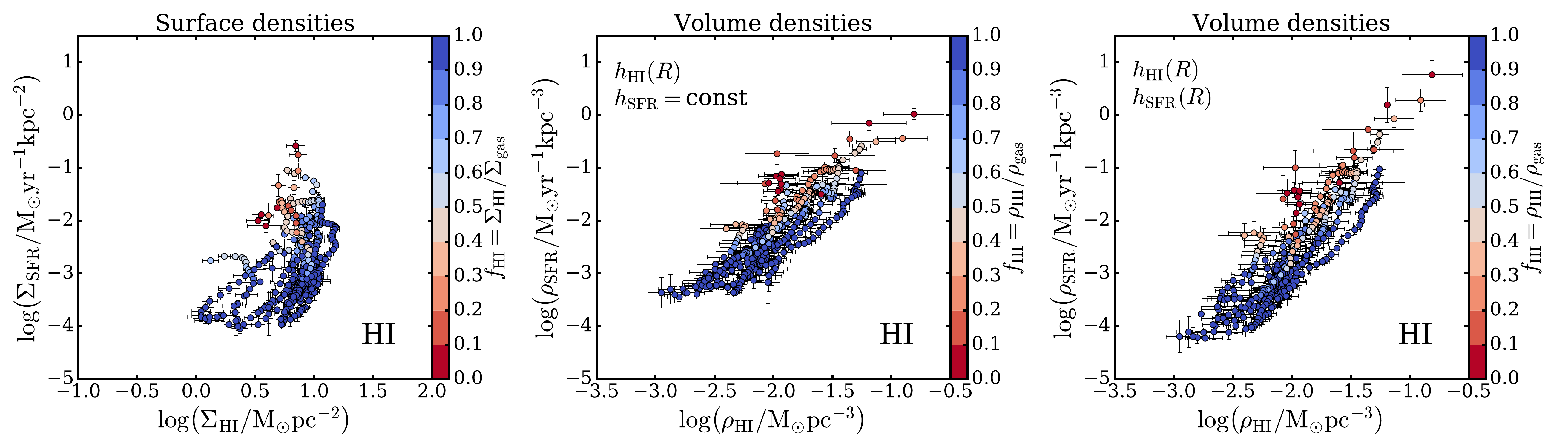}}
\\
\subfloat[][\label{fig:hist_hi}]
{\includegraphics[width=1\textwidth]{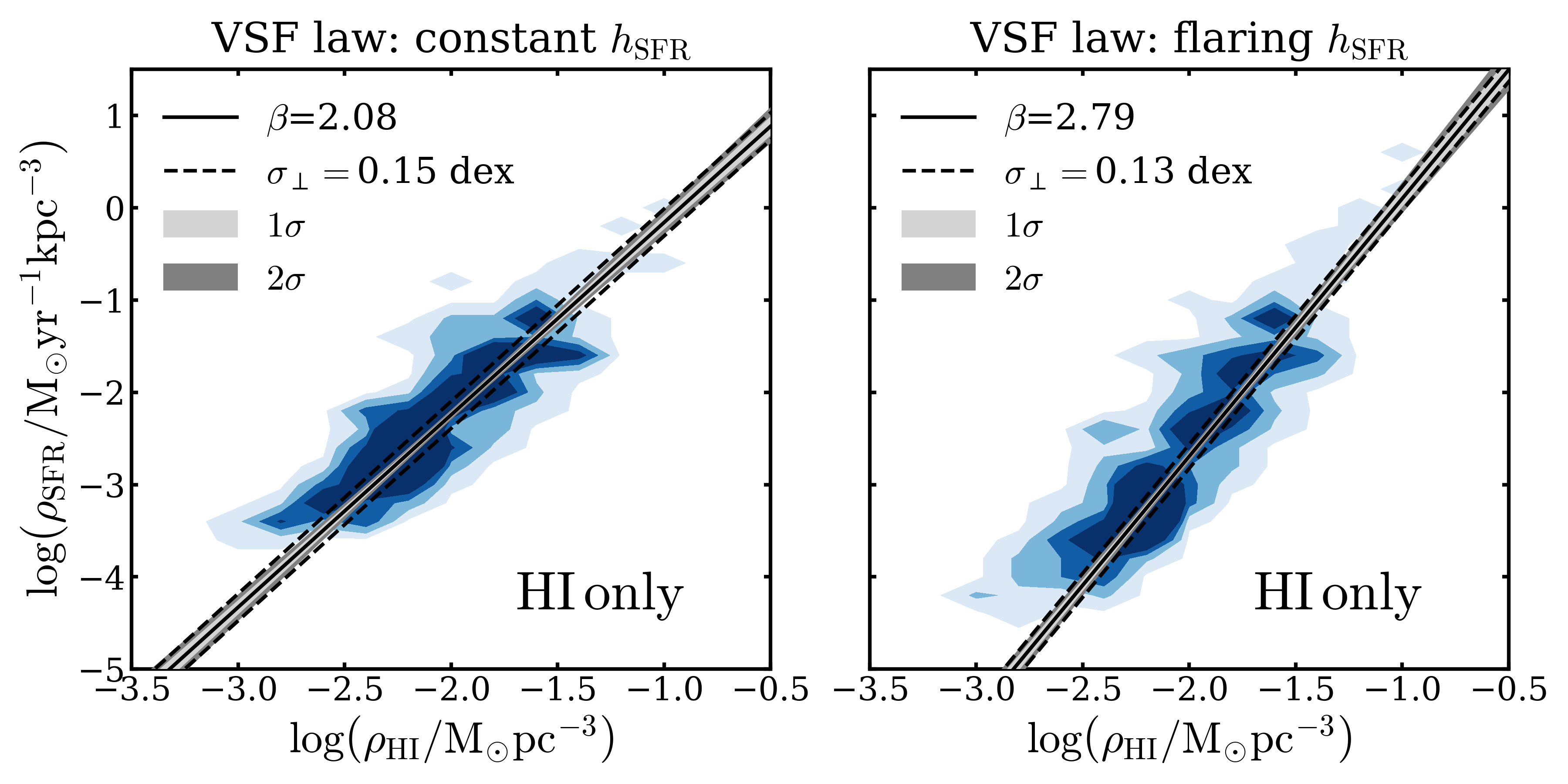}}
\caption{\textit{Upper row:} Same as Fig.~\ref{fig:colorsgals_fgasrho} but with the $x$-axis showing the HI alone surface and volume densities. 
\textit{Lower row:} VSF law between SFR and HI volume densities. See Fig.~\ref{fig:hist_gas} for description.}
\label{fig:fgas_cont_hi}
\end{figure*}

We then investigated if some correlation exists between SFR and gas in the atomic phase. 
In Sec.~\ref{sec:sfr_vol_dens}, the flaring $h_\mathrm{SFR}(R)$ is defined as the weighted mean between $h_\mathrm{HI}$ and $h_\mathrm{H_2}$ according to the 
gas fractions.
Given that only the atomic gas is considered in this case, the SFR flaring scale height is assumed equal to the HI scale height, while the constant 
$h_\mathrm{SFR}$ remains 100 pc as in Sec.~\ref{sec:totgas_corr}. 
Fig.~\ref{fig:colorsgals_fgasrho_hi} compares the correlations between HI and SFR based on surface or volume densities, the points are colour-coded according 
to the HI fraction with respect to the total amount of gas (as in Fig.~\ref{fig:colorsgals_fgasrho}). 
As expected, we found no correlation in the surface-based panel (left), as one order of magnitude in range of HI surface density corresponds to almost 
four orders of magnitude in range of SFR surface densities. 
On the other hand, a tight correlation emerges using the volume densities. 
The implications of this remarkable result are discussed in Sec.~\ref{sec:discussion}.

To determine the HI VSF law parameters, we followed the same procedure as in Sec.~\ref{sec:totgas_corr}, but we defined the model in the 
MCMC fitting (see Appendix~\ref{ap:posteriors} for details) as
\begin{equation}\label{eq:vsf_law_log_hi}
\log \rho_\mathrm{SFR} = \log B + \beta \log \rho_\mathrm{HI} \, .
\end{equation}
We found the slope and the intrinsic scatter, respectively, between 2.1 and 2.8 and 0.15 dex and 0.13 dex, depending on the choice of $h_\mathrm{SFR}$. 
This result indicates a strong link between SF and the atomic gas, in particular in low density environments, where the HI disc is considerably thick. 
Fig.~\ref{fig:hist_hi} shows the volume density data points as contours and the best-fit relation $\pm \sigma_\perp$ is represented by the solid black line. 
For completeness, we compared this correlation with that obtained with the $h_\mathrm{SFR}(R)$ for the case with total gas 
(instead of $h_\mathrm{SFR}(R)=h_\mathrm{HI}(R)$), finding no significant difference between the results in the two cases.

\subsection{Molecular gas versus star formation rate}\label{sec:h2only_corr}
\begin{figure*}
\centering
\subfloat[][\label{fig:colorsgals_fgasrho_h2}]
{\includegraphics[width=1\textwidth]{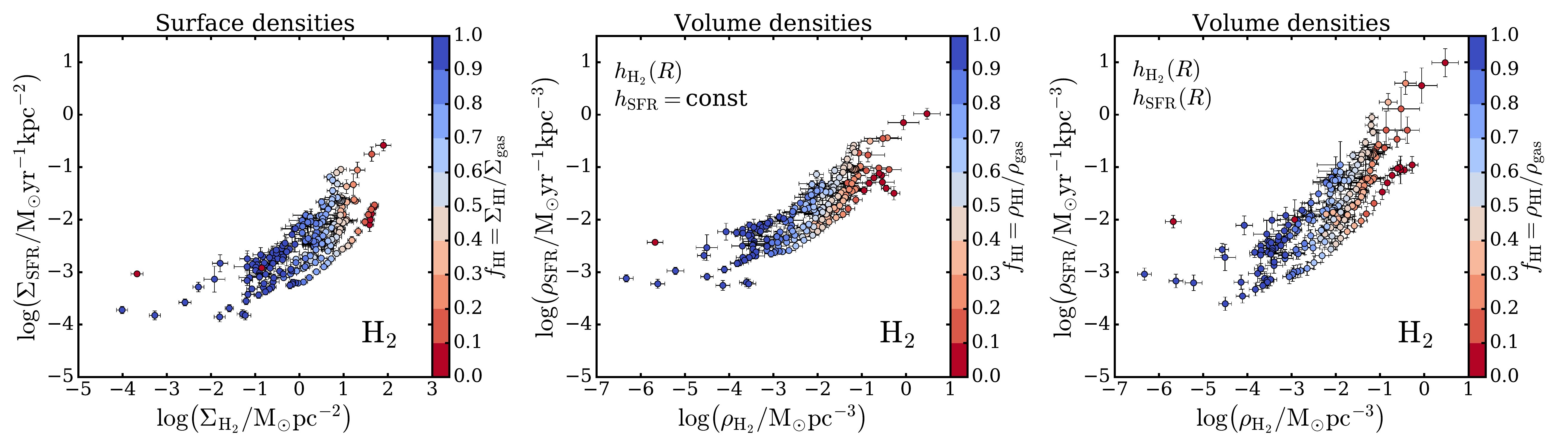}}
\\
\subfloat[][\label{fig:hist_h2}]
{\includegraphics[width=1\textwidth]{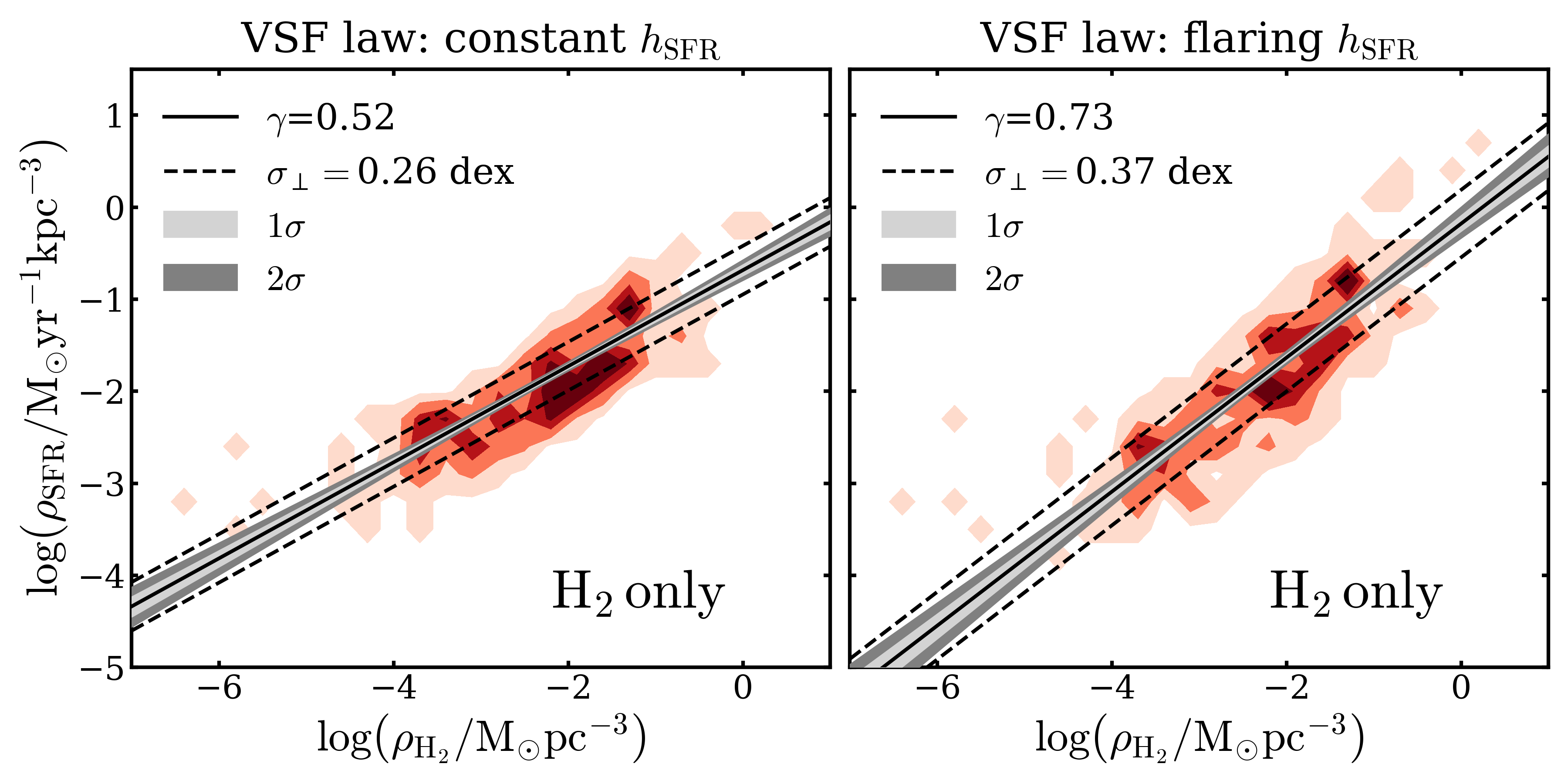}}
\caption{\textit{Upper row:} Same as Fig.~\ref{fig:colorsgals_fgasrho} but with the $x$-axis showing the H$_2$ alone surface and volume densities. 
\textit{Lower row:} VSF law between SFR and H$_2$ volume densities. See Fig.~\ref{fig:hist_gas} for description.}
\label{fig:fgas_cont_h2}
\end{figure*}

Let us now focus on the correlation between the SFR and the molecular gas phase. 
In this case the flaring SFR scale height is equal to $h_\mathrm{H_2}$. 
Fig.~\ref{fig:colorsgals_fgasrho_h2} compares the correlations between H$_2$ and SFR surface and volume densities with the points coloured 
according to the HI fraction with respect to the total amount of gas. 
As expected, there is a clear sign of some H$_2$-SFR correlation in all the three panels, but the volumetric relations appear to be more scattered than both 
the total gas-SFR and the HI-SFR VSF laws. 
In addition, it seems that the molecular correlation is no more valid in the low density regime or that it is not a single power law. 
Indeed, there are hints of a bend both in the surface and volume density plots located at about 1 M$_\odot$pc$^{-2}$ and 0.01 M$_\odot$pc$^{-3}$, 
respectively, where the environment is no more H$_2$-dominated (see Sec.~\ref{sec:discussion} for discussion). 

Again, we performed an MCMC fitting to determine the parameters of the H$_2$-SFR VSF law, which was modelled as
\begin{equation}\label{eq:vsf_law_log_h2}
\log \rho_\mathrm{SFR} = \log \Gamma + \gamma \log \rho_\mathrm{H_2} \, .
\end{equation}
We found that the slope is between 0.5 and 0.7 but, in this case, the intrinsic scatter is 0.3-0.4 dex, so two times larger than the previous cases with the 
total and atomic gas. 
Fig.~\ref{fig:hist_h2} shows the volume density data points as contours and the best-fit relation as the solid black line. 
As in Sec.~\ref{sec:hionly_corr}, we tested the case with the $h_\mathrm{SFR}(R)$ for total gas (instead of $h_\mathrm{SFR}(R)=h_\mathrm{H_2}(R)$) and found 
no significant difference between the results. 

We could argue that the molecular gas VSF law may be sensitive to the possible underestimate of the $\alpha_\mathrm{CO}$ factor \citep[see][]{2013Sandstrom}, as 
there are fewer $\rho_\mathrm{H_2}$ points than those of $\rho_\mathrm{gas}$. 
However, given the scatter of the relation in Fig.~\ref{fig:hist_h2}, it is unlikely that shifting rightward 25 points out of a total of 249 could affect the 
best-fit parameters significantly. 
As further test of the influence of $\alpha_\mathrm{CO}$ on the VSF laws parameters, we repeated the entire procedure and the MCMC fit using the molecular gas 
surface densities of \cite{2008Leroy}. 
We found $\gamma = 0.60 \pm 0.03$ and $\gamma = 0.95 \pm 0.03$ with the constant $h_\mathrm{SFR}$ and $h_\mathrm{SFR}(R),$ respectively, and 
$\sigma_\perp = 0.60 \pm 0.03$ in both cases, which is fully compatible within the uncertainties with the relation shown in Fig.~\ref{fig:hist_h2}. 

To quantitatively compare the molecular VSF law with its surface-based version \citep[e.g.][]{2008Bigiel}, we used an MCMC fitting on the H$_2$ surface 
densities leaving the slope, the $y$-intercept and $\sigma_\perp$ free. 
The resulting best-fit relation is flatter than a linear correlation; indeed the slope is 0.62. 
However $\sigma_\perp$ is 0.3, so the molecular surface-based relation is more scattered than both the total gas and the HI VSF laws. 
Probably, the reason why we find a different slope with respect to the unitary value estimated by some authors \citep[e.g.][]{2008Bigiel} is the 
$\alpha_\mathrm{CO}$ factor, as the linear relation is usually obtained using the MW $\alpha_\mathrm{CO}$ factor for all the galaxies. 
If we fix the slope to 1 and repeat the fit, the resulting scatter is even larger, about 0.4 dex. 
The latter is compatible with the intrinsic scatter of the molecular VSF law (0.25-0.37 dex), thus the volume-based approach does not improve the scatter 
significantly with respect to the surface-based relation. 
Indeed, the molecular gas distribution typically extends to smaller radii with respect to the atomic gas distribution (Fig.~\ref{fig:plot_scaleheights}). 
This means that the scale height at the outermost radius where the H$_2$ is measured is only slightly higher than that at the innermost radii. 
Hence, the conversion to volume densities through the scale height has typically a much milder effect on the molecular gas distribution with respect to the atomic gas. 
Moreover, the $\alpha_\mathrm{CO}$ factor is a further source of scatter in the H$_2$-SFR relation which cannot be reduced by our method. 
\renewcommand{\arraystretch}{1.5}
\begin{table}
        \centering
        \caption{MCMC best-fit parameters of VSF laws. 
        The first and second columns report the gas phases and the SFR scale height involved. 
        The other columns provide best-fit slope, orthogonal intrinsic scatter $\sigma_\perp$, and $y$-intercept with their uncertainties.}
        \label{tab:mcmc_results}
        \begin{tabular}{ll|cccc}
        \hline\hline

        Gas     & $h_\mathrm{SFR}$      & Slope                 & $\sigma_\perp$         &  $y$-intercept                                \\
                &                       &                       &(dex)                  & $\left( \log \frac{\rho_\mathrm{SFR}}{\mathrm{M}_\odot \mathrm{yr}^{-1} \mathrm{kpc}^{-3} } \right)$              \\
        \hline
        HI+H$_2$& constant              & 1.34$^{+0.02}_{-0.02}$&0.13$^{+0.01}_{-0.01}$& 0.23$^{+0.01}_{-0.01}$\\
        HI+H$_2$& flaring               & 1.91$^{+0.03}_{-0.03}$&0.12$^{+0.01}_{-0.01}$& 0.90$^{+0.02}_{-0.02}$\\
        HI      & constant              & 2.09$^{+0.06}_{-0.06}$&0.15$^{+0.01}_{-0.01}$& 1.93$^{+0.02}_{-0.02}$\\
        HI      & flaring               & 2.79$^{+0.08}_{-0.08}$&0.13$^{+0.01}_{-0.01}$& 2.89$^{+0.03}_{-0.02}$\\
        H$_2$   & constant              & 0.52$^{+0.02}_{-0.02}$&0.26$^{+0.02}_{-0.01}$& -0.68$^{+0.02}_{-0.02}$\\
        H$_2$   & flaring               & 0.73$^{+0.03}_{-0.03}$&0.37$^{+0.02}_{-0.02}$& -0.18$^{+0.03}_{-0.03}$\\
        \hline
\end{tabular}
\end{table}

%
\section{Discussion}\label{sec:discussion}
\subsection{Comparison with other works on gas and star scale heights}
The thickness of gaseous and stellar discs in galaxies has been studied for years using both theoretical and observational approaches. For example, our method is very similar to that implemented by \cite{2011Banerjee}. 
They calculated the atomic gas scale heights for DDO 154 and IC 2574 through an iterative algorithm based on the hydrostatic equilibrium. 
The parametric mass models of DDO 154 and IC 2574 were taken from \cite{2008Deblok} and \cite{2008Oh}, thus the first is equal to the model in 
Table~\ref{tab:gravpot_params_main} and the second does not differ significantly. 
These works adopted the velocity dispersion measured by \cite{2009Tamburro} using the 2D method on THINGS data cubes. 
In order to directly compare the scale heights, we must verify that \cite{2011Banerjee} velocity dispersion is the same as we measured. 
For DDO 154, they assumed $\sigma_\mathrm{HI} = 8$ \kms and constant with radius, so that their outermost scale height ($R \approx$ 6 kpc) is about 1 kpc. 
Within uncertainties, their result is compatible with our scale height at 6 kpc, which is 1.4 kpc as our velocity dispersion is about 10 \kms (25\%) higher. 
For IC 2574, the declining radial profile of the velocity dispersion assumed by \cite{2011Banerjee} is 20\% lower everywhere than that shown in 
Fig.~\ref{fig:vdisp_fits}. 
The resulting scale height profiles are perfectly compatible within the errors. 

Recently, \cite{2018Patra} estimated the scale heights of the HI and H$_2$ discs in hydrostatic equilibrium for NGC 7331. 
The gravitational potential model was taken from \cite{2008Deblok}, therefore it is similar to that reported in Table~\ref{tab:gravpot_params_main}. 
There are two differences from our method. 
The first is that the \cite{2018Patra} stellar disc was assumed to be in \textquotedblleft hydrostatic equilibrium\textquotedblright, so the stellar 
scale height was derived iteratively as for the gas components. 
The second difference is that the author assumed the velocity dispersion to be constant with radius. 
As a consequence, the comparison is not straightforward but, for $\sigma_\mathrm{HI} =$ 10 \kms, \citeauthor{2018Patra} found $h_\mathrm{HI} \approx$ 
230 pc at $R=10$ kpc, which is compatible with our result of about 400 pc at same radius but with $\sigma_\mathrm{HI} =$ 20 \kms. 

None of the galaxies in our sample are edge-on, so an accurate direct measurement of the gas disc thickness is not possible. 
However, there are several observational clues that gas discs flares with radius; we give just two recent examples. 
\cite{2011Yim,2014Yim} adopted the method proposed by \cite{1996Olling} to measure simultaneously the inclination of the galaxy and the thickness of stellar and gaseous discs.  
For four star-forming nearby galaxies, they found that both the thicknesses of the atomic gas disc \citep[see also][]{2017Peters} and stellar disc 
flare with radius. 
They also derived the thickness of the CO disc finding clear signs of the flaring for NGC\,891, while the CO flare was not clearly visible for the other 
galaxies, given the larger uncertainties. 

In the MW, \cite{2017Marasco} measured the scale height of HI and H$_2$ vertical distributions and found that the HI scale height increases of a factor 2 
from about 100 pc at $R \approx 2.5$ kpc to about 200 pc in the solar neighbourhood. 
The molecular gas scale height was instead found to be nearly constant with radius, given the large associated uncertainties. 

Concerning the SFR scale height, if our assumption is correct and the scale height of SFR flares with radius, we should observe that the thickeness of 
the disc of young stellar populations in galaxies increases with radius, at least in the outskirts. 
\cite{2017Mackereth} used SDSS-APOGEE survey data to decompose the MW stellar disc according to age, metallicity, and 
[$\alpha$/Fe], and analysed the radial and vertical distributions the different populations. 
They found that the scale height of young populations, which have mainly high metallicity and low [$\alpha$/Fe], flares with radius 
\citep[see also][]{2018Xiang}. 
On the other hand, the scale height of old stars, which tend to have high [$\alpha$/Fe] and low metallicity, is higher but radially flatter than that of the 
young populations. 

\subsection{Comparison with other works on volume-based SF laws}
Other authors investigated the existence of a volumetric relation. 
The first VSF law was proposed by \cite{1959Schmidt}, who linked the HI and SFR through a single power law with slope between 2 and 3. 
Hence, we should compare his result with our VSF law with HI alone. 
Interestingly, the best-fit slope found in Sec.~\ref{sec:hionly_corr} is perfectly compatible with Schmidt's estimate, suggesting the existence of a 
universal correlation involving the atomic phase of gas. 

\cite{2008Abramova} is probably the work most similar to ours. 
For a sample of seven galaxies (including the MW), they calculated the HI and the H$_2$ scale heights assuming hydrostatic equilibrium 
\citep[see also][]{2002NarayanJog} and then converted the azimuthally averaged surface density radial profiles to volume densities. 
For the SFR scale height, they tried two approaches: one assumed a constant scale height and the other used the stellar disc thickness. 
However, they neglected the radial decrease of the velocity dispersion for the gas components, which is the most significant difference with our approach. 
They found that the gas and SFR volume densities are better correlated with respect to the surface densities, but the slope of the volumetric relation 
for their galaxies has large variations between 0.8 and 2.4, which are on average close to 1.5. 
They drew the same result for the molecular gas alone with average slope was close to 1. 
Given the significant difference between the slope of SF laws for single galaxies, they concluded there was an absence of a universal relation. 
This is probably due to the assumption of radially constant velocity dispersion profiles; our result is however in agreement within the uncertainties 
with their average slope of 1.5. 

More recently, \cite{2012Barnes} studied the link between gas and SFR in the outer disc of seven nearby star-forming galaxies. 
They found a very steep surface-based SF law in the form of a single power law with index $2.8 \pm 0.3$. 
Then, they estimated the HI disc thickness through the full width at half maximum (FWHM) of the gas vertical profile 
\citep[Eq. 20 in][]{2011vanderKruitFreeman}, finding that this thickness flares with radius. 
They used this proxy to convert the total gas surface densities radial profiles to volume densities and assumed 100 pc as fiducial value for the 
SFR scale height. 
They found a volume-based correlation with index $1.5 \pm 0.1$ between gas and SFR, which is in agreement with our result (see \citealt{1991Tenjes} for a similar 
study on M31). 
 
Concerning Galactic studies, \cite{2017Sofue} used 3D maps of HI from the LAB survey \citep{2005Kalberla} and H$_2$ from the CO survey \citep{2001Dame} 
to estimate the gas volume densities out to 20 kpc. 
\citeauthor{2017Sofue} measured the SFR volume density from the HII region catalogue and investigated the existence of volumetric correlations with 
total gas, HI only, and H$_2$ only. 
The author used two approaches: the first consists in dividing the data in radial bins, while the second considers the whole radial range (0-20 kpc). 
In the first case, they found that any VSF law showed radial variations both in the index and the normalisation. 
On the other hand, the second method revealed a correlation with index of $2.01 \pm 0.02$ for the VSF law with total gas, while the relations involving 
the molecular and the atomic gas only were found to have a slope of $0.70 \pm 0.07$ and of $2.29 \pm 0.03,$ respectively. 
These results are in excellent agreement with our findings. 

\cite{2012Krumholz} formulated a theory involving a molecular and volumetric SF law and compared it to the observed correlations. 
They gathered a collection of the correlations between gas and SFR using both resolved observations of MW molecular clouds and Local galaxies, and
unresolved observations of local discs and high redshift starbursts. 
These authors explained the diversity of the observed gas-SFR correlations as the result of the variety of three-dimentional sizes and internal clumpiness, as the 
volume of the observed region can be very different at fixed surface density. 
Hence, they removed these projection effects by calculating the free-fall time specifically for each different regime, from molecular clouds to high 
redshift galaxies, and found that all the data fall on a single power-law relation. 
In other words, they did not convert the surface densities to volume densities, but they instead built the free-fall timescale using a different 
prescription for molecular clouds, disc galaxies, and starbursts, obtaining a correlation between $\Sigma_\mathrm{SFR}$ and 
$\Sigma_\mathrm{gas} / t_\mathrm{ff}$. 
However, the approach of \cite{2012Krumholz} differs from ours in many aspects. 
For example, they assumed that the star-forming gas is exclusively molecular, so the free-fall time is always calculated for the molecular phase. 
In addition, they did not take into account the vertical hydrostatic equilibrium for the gas and neglected the scale height flaring with radius. 

\subsection{Physical interpretation}
We conclude by discussing some potential physical interpretations of our findings, starting from the most straightforward. 
In order to form stars, the gas must be cold and dynamically unstable, therefore the SF timescale is given by the longest between the dynamical and 
cooling timescales \citep[see e.g.][]{2007CiottiOstriker}. 
A key result of our investigation is the superlinear correlation between the SFR and total gas volume densities in the form of a single power law. 
If we believe that the true SFR scale height is bracketed between the constant and the flaring profiles, the index of the VSF law with total gas should 
reasonably be between the best-fits slopes of 1.3 and 1.9. 
If the index is 1.5, then the physical explanation of the correlation may come from the gravitational instability of the gas, indicating that the cooling 
timescale is shorter than the dynamical timescale. 
Hence, $\rho_\mathrm{SFR} \propto \rho_\mathrm{gas}/\tau_\mathrm{ff} \propto \rho_\mathrm{gas}^{3/2}$ \citep[e.g.][]{1977Madore,2006Li}. 
On the other hand, it is well known that the interstellar medium is not a continuous fluid, but it is mostly in gas clouds and filaments, 
therefore this interpretation of the global correlation may be not suitable to describe SF on the scales of single clouds. 
However, our results appear to indicate that the average SFR density at different locations in a galaxy disc is rather precisely regulated by the total volume 
density of the locally available gas. 

Moreover, the fact that the observed break in the Kennicutt law disappears after the conversion to volume densities indicates that it is probably caused 
by the flaring of the gas disc. 
This was also suggested by \cite{2015Elmegreen}, who aimed to explain the change in the index of the surface-based SF laws. 
He showed that the classical Kennicutt law between total gas and SFR surface densities is valid in the main regions of spiral galaxies, where he assumed 
that the scale height is almost constant. 
In the outskirts instead, he found a steeper index of 2 for the surface-based law, as the gas disc thickness increases with radius. 
However, the DM contribution was not included in the model of the galactic gravitational potential, thus the gas is completely self-gravitating 
in the outer regions and the resulting gas scale height is overestimated. 

The tight correlation between the atomic gas and the SFR is the most surprising result of our work. 
In this case, the interpretation is more difficult and uncertain. 
If the molecular gas is the prerequisite for SF, why should we observe a correlation between HI and SFR? 
It is well known that the molecular gas forms from atomic gas, so the possible explanation for the HI VSF law is that the atomic gas is a good tracer of 
the cold (and molecular) star-forming gas both in low density and, to some extent, high density regions. 
Indeed, the outskirts of spiral galaxies and dwarf galaxies are often metal poor and low density environments, hence the amount of CO is probably too low 
to be detected. 
This scenario could explain the observed extended UV discs (XUV) \citep{2007Thilkera,2007Thilkerb}, showing that SF can occur also in the 
outermost and HI-dominated regions of disc galaxies \citep[see also][]{1998Ferguson}, where the metallicity is expected to be very low. 

Taken to extremes, the HI VSF law could also mean that molecular gas is not always a prerequisite for SF and the atomic gas plays a key role in the process. 
\cite{2012KrumholzHI} showed that SF can occur in cold atomic gas (at extremely low metallicity) rather than in molecular gas, 
thanks to the efficient cooling by C$^+$. 
In such peculiar conditions, the timescale to convert HI to H$_2$ is longer than the timescales to reach the thermal equilibrium (cooling time) 
and gravitational collapse (free-fall time). 
Hence, atomic gas can efficiently cool and form stars, but it does not have enough time to turn into a significant amount of H$_{2}$. 
Similarly, \cite{2012GloverClark} investigated whether or not the molecular gas is essential for SF. 
They performed a set of numerical simulations of dense clouds using different chemical prescriptions: one in which the gas remains atomic for the whole 
cloud evolution, a second including H$_2$ formation, and a third following both H$_2$ and CO formation. 
They found that the SF process is very similar in all the simulations and concluded that the molecular gas is not a prerequisite for SF, as the gas can 
efficiently cool thanks to C$^+$ line emission at low density and  by energy transfer from gas to dust at high density. 
On the contrary, they found that including or not the dust shielding is fundamental, as it allows the gas to cool below 100 K and form stars. 
In other words, the ability of clouds to shield themselves from the interstellar radiation field is the key to SF. 
Moreover, the authors concluded that the observed correlation between the molecular gas and SFR surface densities originates from the fact that both 
the terms correlate with a third factor, which is the clouds ability to self-shield \citep[see also][]{2007Krumholz}. 

Concerning the H$_2$ VSF law, the interpretation is even more difficult, as the estimate of the molecular gas volume density is problematic. 
This correlation seems to hold for the central parts of galaxies, despite the large uncertainties associated with the 
$\alpha_\mathrm{CO}$ factor, velocity dispersion, and bulge potential. 
Probably, the interplay of these factors causes the large scatter of the molecular VSF law. 
However the surface density law has a similar scatter. 
The molecular gas is mostly in giant molecular clouds, but we are not including any clouds filling factor in our study. 
Thus, the volume density that we calculate is simply a mean value in a region of $\Delta z \approx h_\mathrm{H_2}$ perpendicular to the 
midplane, so our estimate of the volume density is very different from the volume density inside a cloud. 
This could explain why the molecular gas volume densities reach values lower than $10^{-3}$ M$_\odot$ pc$^{-3}$, which corresponds to about 
$10^{-2}$ H$_2$ particles per cm$^{3}$. 

It is interesting to compare our results with the recent work by \cite{2018Catinella}. 
They presented the extended GALEX Arecibo SDSS Survey (xGASS), a census of 1179 galaxies selected by stellar mass ($10^9 \mathrm{M}_\odot < 
\mathrm{M}_\star < 10^{11.5} \mathrm{M}_\odot$) and redshift ($0.01 < z < 0.05$). 
They measured stellar masses, SFRs, and HI masses for all the galaxies and H$_2$ masses for 532 galaxies. 
They found that the gas reservoir in galaxies is on average HI-dominated, while the ratio of the HI to H$_2$ masses slightly increases with increasing 
stellar mass. 
Moreover, for the whole mass range, the HI mass tightly correlates with the dust-unobscured SFR traced by near-ultraviolet\texttwelveudash$r$ colour. 
In light of these results, our correlation between HI and SFR volume densities is not surprising. 

The tight VSF law between total gas and SFR corroborates the idea that the whole gas, including the atomic phase, traces SF in galaxies. 
Then, the HI and H$_2$ VSF laws could help in understanding the mechanism of the conversion of atomic gas to molecular gas and how important this is in the 
whole SF process. 

%

\section{Summary and conclusions}\label{sec:concl}
We investigated the existence of a fundamental SF law based on volume densities of gas and SFR. 
We built VSF laws using the volume densities radial profiles calculated from the surface densities profiles of 12 nearby galaxies. 
To make the conversion to volume densities possible, we assumed the hydrostatic equilibrium and calculated the HI, H$_2$, and SFR scale heights, which required 
two preliminary steps: the first to calculate the total gravitational potential and the second to measure the gas velocity dispersion. 
Using volume densities, we found a correlation between the total gas (HI+H$_2$) and the SFR, which is less scattered than the classical surface-based law. 
Moreover, an unexpected and tight relation between HI and SFR volume densities was discovered, suggesting a profound link between the atomic phase of 
gas in galaxies and SF. 
The H$_2$-only version of the VSF law was found to have a larger scatter with respect to the HI-only and total gas relations, it seems to break down 
in low density and HI-dominated environments. 

Hence, our conclusions are the following. 
\begin{enumerate}
 \item The thickness of gas discs in hydrostatic equilibrium shows a significant flaring with radius, regardless of the galaxy type. 
 This means that assuming a constant scale height for gaseous discs is not a good approximation. 
 \item The total gas and the SFR volume densities are linked by a tight and single power law with index between 1.3 and 1.9, depending on whether a 
 flare in the SFR scale height itself is taken into account or not.  
 \item The break observed in the Kennicutt law may not be indicative of a low SF efficiency of atomic gas at low surface density, but rather be 
 a consequence of the radial flaring of the gas discs.
 \item The SFR volume density also correlates with the HI alone volume density through a single power law with small scatter and index between 2.1 and 2.8. 
\end{enumerate}
The VSF law is likely more fundamental and general than surface-based laws, as it takes into account the three-dimensional distribution of gas and SFR. 
The unexpected and tight correlation between HI and SFR volume densities may be important to unveil the mechanisms that regulate the conversion of gas 
into stars, in particular in low density and HI-dominated environments as dwarf galaxies and the outskirts of spiral galaxies.

\section*{Acknowledgements}
C. B. is grateful to A. Marasco, L. Posti, M. Nori, V. Ghirardini, and E. di Teodoro for inspiring conversations and advice, 
and to Bradley Frank for sharing the surfaced density profiles of molecular gas. 
G. I. is supported by the Royal Society Newton International Fellowship. 
G. P. acknowledges support by the Swiss National Science Foundation, grant PP00P2\_163824. \textquotedblright

%

\bibliographystyle{aa}
\bibliography{paty.bib}

%

\appendix

\section{VSF law using the analytical approximation for the scale height}\label{ap:h_analyt}
In Sec.~\ref{sec:hdefapprox}, we reported an analytic approximation (Eq.~\ref{eq:hdef}) for the scale height of the vertical distribution of a gas disc in hydrostatic 
equilibrium (see \citealt{1990PhDTRomeo,1992Romeo} for a rigorous analytic study of the vertical structure of galactic discs). 
We now show that the results obtained through this definition are compatible with what we have found using the numerical (and more accurate) method. 
Through the Poisson equation of gravity in the $z$ direction, we know that
\begin{equation}\label{eq:poisson}
 \frac{\partial^2 \Phi(R,z)}{\partial z^2} = 4 \pi G \rho (R,z) - 
 \frac{1}{R} \frac{\partial}{\partial R} \left( R \frac{\partial \Phi(R,z)}{\partial R} \right) \, .
\end{equation}
In the midplane, $\frac{\partial \Phi(R,0)}{\partial R} = \frac{V_\mathrm{c}(R)^2}{R}$, where $V_\mathrm{c}$ is the galaxy circular velocity. 
Hence, the r.h.s. of Eq.~\ref{eq:poisson} is
\begin{equation}\label{eq:poisson_rhorot}
 \frac{\partial^2 \Phi(R,z)}{\partial z^2} \approx 4 \pi G \left[ \rho (R,0) + \rho_\mathrm{rot}(R) \right] \, ,
\end{equation}
where the rotational density $\rho_\mathrm{rot}$ is \citep[see][]{1984Bahcall,1984BahcallCasertano,1995Olling}
\begin{equation}\label{eq:rhorot_def}
 \rho_\mathrm{rot}(R) \equiv - \frac{1}{2 \pi G} \frac{V_\mathrm{c}(R)}{R} \frac{\partial V_\mathrm{c}(R)}{\partial R} \, .
\end{equation}
Substituting Eq.~\ref{eq:poisson_rhorot} in Eq.~\ref{eq:hdef}, we find the following simple and generic analytical formulation for the scale height that can be used 
once the mass distribution of the galaxy components are known:
\begin{equation}\label{hgen_analytic}
 h(R) \simeq \frac{\sigma(R)}{\sqrt{4 \pi G [ \rho (R,0) + \rho_\mathrm{rot}(R)]}} \, .
\end{equation}
We note that, with this approximation, the gas self-gravity is not included and we are assuming a cylindrical mass distribution which, at fixed $R$, does 
not vary with $z$ with respect to its value in the midplane. 

\begin{figure*}
    \includegraphics[width=2\columnwidth]{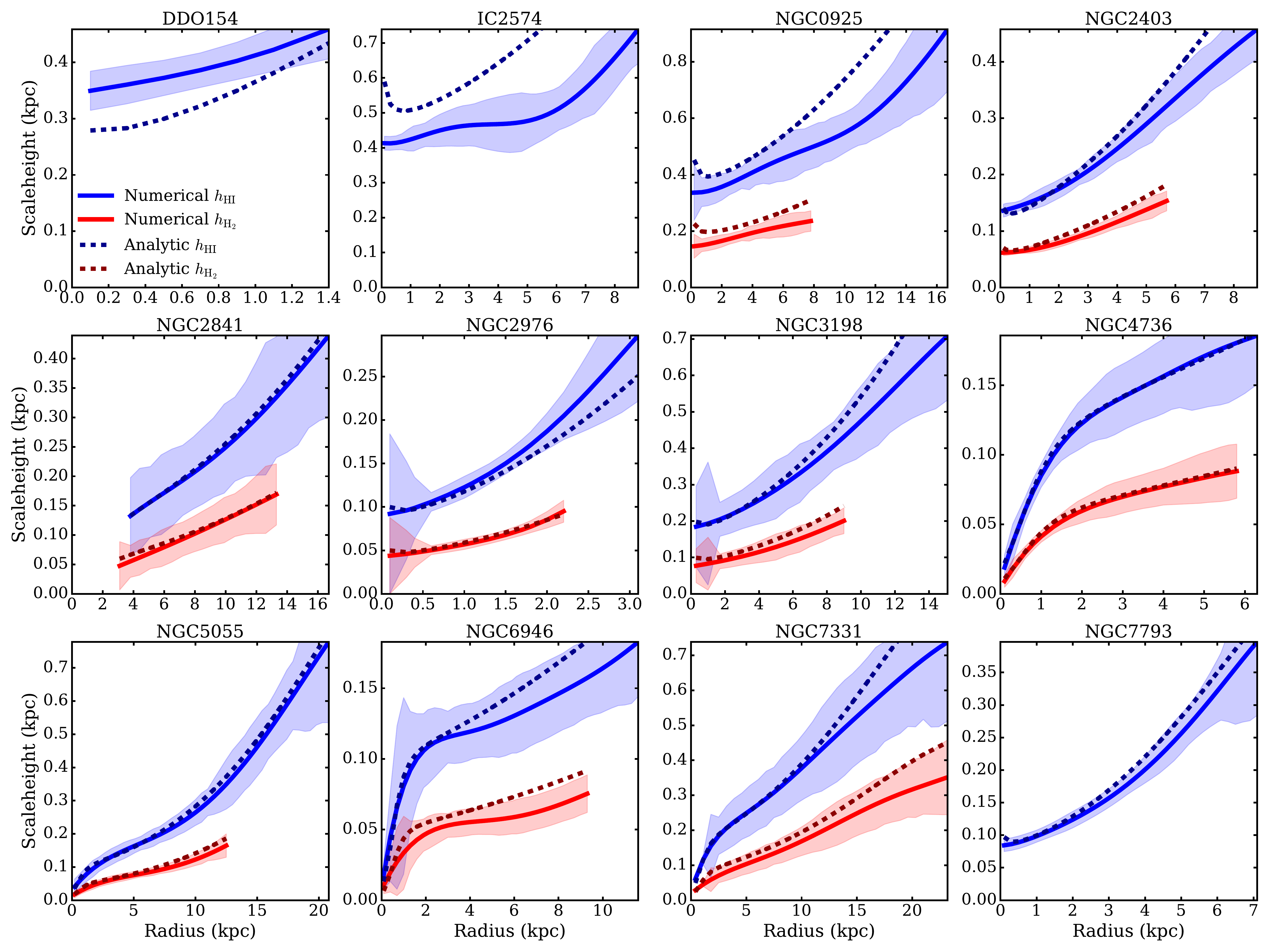}
    \caption{Comparison of the scale heights calculated by Eq.~\ref{hgen_analytic} (dashed line) and by \textsc{Galpynamics} (solid line). 
    The HI and H$_2$ scale heights are in blue and red, respectively.}
    \label{fig:h_anVSnum}
\end{figure*}

For the galaxies in our sample, we assumed that the main mass components are the spherical DM halo and the stellar disc, which are 
modelled by Eq.~\ref{eq:isohalo} or Eq.~\ref{eq:nfwhalo} and Eq.~\ref{eq:expsec_stardisc}, respectively. 
The rotational density for the NFW halo is (with $x=R/R_\mathrm{s}$)
\begin{equation}
 \rho_\mathrm{rot}(x) = - \rho_\mathrm{DM,0} \left[ \frac{x(2x+1)-(1+x^2)\log(x+1)}{(1+x^2)x^3} \right] \,.
\end{equation}
For the isothermal halo, we have (with $y=\sqrt{1+R^2/R_\mathrm{c}^2}$)
\begin{equation}
 \rho_\mathrm{rot}(y) = - \rho_\mathrm{DM,0} \left[ \frac{\arctan(y)(1+y^2)-y}{(1+y^2)y^3} \right] \, .
\end{equation}
The exponential disc circular velocity is given by Eq. 2.165 in \cite{2008BinneyTremaine}, so the rotational density is
\begin{equation}
 \rho_\mathrm{rot}(k) = - \frac{\Sigma_{\star,0} }{16 R_\star} \left[ k A(k/2) + 8 B(k/2) \right] \, ,
\end{equation}
where $k=R/R_\star$, $A(k/2)=3 K_0 I_1  + K_2 I_1  - 3 K_1 I_0 - I_2 K_1$ and $B=I_0 K_0 - I_1 K_1$ being $K_0$, $K_1$, $I_0$, and $I_1$ the 
modified Bessel functions. 
The stellar bulge is modelled as an exponential sphere (Eq.~\ref{eq:bulge_expball}) with circular velocity given by Eq.~\ref{eq:vcirc_expball}, so
\begin{equation}
 \rho_\mathrm{rot}(R) = \frac{V_\mathrm{c}^2}{4 \pi G R^2} - \rho_\mathrm{b}(R) \, .
\end{equation}
Table~\ref{tab:gravpot_params_main} provides all the parameters to calculate the mass distributions and rotational densities. 
Fig.~\ref{fig:h_anVSnum} compares $h_\mathrm{HI}$ and $h_\mathrm{H_2}$ calculated through Eq.~\ref{hgen_analytic} and with \textsc{Galpynamics}, the 
velocity dispersion is modelled as explained in Sec.~\ref{sec:hivdisp} and Sec.~\ref{sec:h2vdisp}. 
For the majority of the galaxies in our sample, the analytical estimate is compatible with the numerical scale height within the uncertainties. 
Hence, the SFR scale height calculated through Eq.~\ref{eq:hSFR_wmean} but using the approximated $h_\mathrm{HI}$ and $h_\mathrm{H_2}$ is approximately 
equivalent to that shown in Fig.\ref{fig:plot_scaleheights}. 

Then, we use the analytical scale heights to convert the observed surface densities (see Sec.~\ref{sec:sample_surf}) to volume densities through 
Eq.~\ref{eq:midplane}. 
We have therefore all the elements to build the VSF laws and perform an MCMC fitting to derive slope, $y$-intercept and scatter of the VSF laws 
with total gas, HI only, and H$_2$ only. 
The results are reported in Table~\ref{tab:mcmc_results_an} and are compatible within the uncertainties with those in Table~\ref{tab:mcmc_results}. 

\renewcommand{\arraystretch}{1.5}
\begin{table}
        \centering
        \caption{MCMC best-fit parameters for the VSF laws in the analytical approximation for the scale heights. 
        The first and second columns report the gas phases and the SFR scale height involved. 
        The other columns provide best-fit slope, orthogonal intrinsic scatter $\sigma_\perp$, and $y$-intercept with their uncertainties.}
        \label{tab:mcmc_results_an}
        \begin{tabular}{ll|cccc}
        \hline\hline
        Gas     & $h_\mathrm{SFR}$      & Slope                 & $\sigma_\perp$         &  $y$-intercept                                \\
                &                       &                       &(dex)                  & $\left( \log \frac{\rho_\mathrm{SFR}}{\mathrm{M}_\odot \mathrm{yr}^{-1} \mathrm{kpc}^{-3} } \right)$              \\
        \hline
        HI+H$_2$& constant              & 1.31$^{+0.03}_{-0.02}$&0.13$^{+0.01}_{-0.01}$& 0.25$^{+0.02}_{-0.02}$\\
        HI+H$_2$& flaring               & 1.91$^{+0.03}_{-0.03}$&0.11$^{+0.01}_{-0.01}$& 0.94$^{+0.02}_{-0.02}$\\
        HI      & constant              & 1.97$^{+0.06}_{-0.06}$&0.15$^{+0.01}_{-0.01}$& 1.79$^{+0.02}_{-0.02}$\\
        HI      & flaring               & 2.70$^{+0.07}_{-0.07}$&0.13$^{+0.01}_{-0.01}$& 2.76$^{+0.02}_{-0.02}$\\
        H$_2$   & constant              & 0.52$^{+0.02}_{-0.02}$&0.27$^{+0.02}_{-0.01}$& -0.66$^{+0.02}_{-0.02}$\\
        H$_2$   & flaring               & 0.73$^{+0.03}_{-0.03}$&0.37$^{+0.02}_{-0.02}$& -0.18$^{+0.03}_{-0.03}$\\
        \hline
\end{tabular}
\end{table}

\section{\textsc{$^{\mathrm{3D}}$Barolo} set-up}\label{ap:3DB}
As mentioned in Sec.~\ref{sec:hivdisp}, we derived the HI velocity dispersion using 3DB on publicly available 21 cm data cubes from the survey THINGS. 
In a broad outline, the tilted-ring modelling consists in decomposing the rotating disc of a galaxy into a series of circular rings with radius $R$ and 
characterised by kinematic and geometrical parameters. 
For each sampling radius, 3DB first builds a ring model described by inclination, position angle, and rotation velocity, then the model is compared to real 
data and the parameters of the ring are updated until the residuals between the model and observations are minimised. 
Before the comparison, the model is convolved with the point spread function (PSF) by degrading it to the same spatial resolution of observations. 
This step is fundamental to account for the beam smearing, which can affect the determination of the velocity dispersion. 
Each ring is fully described by the following parameters: the spatial coordinates of the centre $(x_0,y_0)$, systemic velocity $V_\mathrm{sys}$, 
inclination $i$, position angle $\phi$, rotation velocity $V_\mathrm{rot}$,  velocity dispersion $\sigma_\mathrm{HI}$, face-on HI 
column density $\Sigma_\mathrm{HI}$, and scale height of the gas layer $z_0$. 

We used the robust weighted data cubes to ensure a reliable measurement of the line broadening due to the gas turbulence. 
Indeed, \cite{2017Ianjamasimanana} showed that the shape of the beam of natural weighted data cubes significantly differs from a Gaussian, causing 
an overestimate of the velocity dispersion, while the robust weighted data cubes are less affected by this bias. 
\cite{2008Leroy} measured the surface densities of gas and SFR at resolutions of 400 pc and 800 pc for dwarf galaxies and normal spirals, respectively. 
For all the galaxies, we aimed to have a common spatial resolution that is not only compatible with their surface density measurements, but also 
as high as possible to preserve a good sampling of the velocity dispersion radial trend. 
The most distant galaxy, NGC 2841, was observed at about 400 pc of spatial resolution, and consequently this is our upper limit for homogeneous resolution. 
Hence, we smoothed all data cubes to the same spatial resolution of about 400 pc to improve the S/N in the data cubes (see Table~\ref{tab:bb_params} 
for corresponding beam size). 
Secondarily, this resolution is compatible with the drift scale, which is the displacement between a young star and its parent cloud \citep[see][]{2008Koda}. 

In order to set 3DB initial parameters, we made the following assumptions.
\begin{itemize}
 \item 
 \textit{HI column density}: $\Sigma_\mathrm{HI}$ was removed from the list of free parameters choosing one of the two 3DB normalisation methods. 
 It is possible to normalise the model flux to the observed intensity map by a pixel-by-pixel (local) or azimuthal comparison. 
 The local normalisation better represents the non-axisymmetric features and prevents under-dense or over-dense regions from influencing 
 the residuals minimisation. 
 However, the weakness of this choice is that, in some cases, the algorithm is not able to reliably estimate the radial variation of the inclination and thus the output profile for the inclination varies unrealistically. 
 Therefore, it is advisable to set the inclination  to a fixed value when using the local normalisation. 
 For this reason, when the S/N in the 400 pc resolution data cube is low (as for the most distant galaxies, NGC 2841, NGC 3198, and NGC 7331), we choose 
 the azimuthal normalisation. 
 The normalisation for each galaxy is reported in Table~\ref{tab:bb_params} (L=local, A=azimuthal).  
 In the vast majority of cases, the assumption on the normalisation does not affected the fit nor the dispersion velocity measurements. 
 
 \item 
 \textit{Scale height}: 3DB is insensitive to the scale height as the tilted ring fitting procedure is done ring-by-ring, while for thick discs 
 one line of sight can intersect emission from different annuli because of the projection effects of inclination. 
 \cite{2017Iorio} found that assuming a constant scale height does not significantly affect the estimate of the kinematical parameters 
 in their sample of dwarf galaxies or, at least, these differences are compatible with the errors. 
 The galaxies in our sample are more massive than \cite{2017Iorio} dwarfs, so the thickness bias is expected to be even milder for our galaxies. 
 Following \cite{2017Iorio}, we adopted $z_0=100$ pc, which is smaller than the spatial resolution and constant for each ring. 
 
 \item 
 \textit{Systemic velocity}: Before fixing $V_\mathrm{sys}$, we compared the values reported in \cite{2008Deblok} with the systemic velocity obtained 
 from the global line profile by $V_\mathrm{sys}=0.5 \left( \mathrm{V}^{20\%}_\mathrm{app} + \mathrm{V}^{20\%}_\mathrm{rec} \right)$, where 
 $\mathrm{V}^{20\%}_\mathrm{app}$ and  $\mathrm{V}^{20\%}_\mathrm{rec}$ are the velocities corresponding to the 20\% of flux peak for the approaching and 
 receding sides of the galaxy. 
 The results are compatible for all the galaxies except IC 2574. 
 In this case, we found $V_\mathrm{sys}=44.9$ \kms, which is lower than the 53.1 \kms estimate by \cite{2008Deblok} but compatible with the measurement with 
 3DB, so we chose the former.  
 
 \item 
 \textit{Galaxy centre}: $(x_0,y_0)$ is fixed to the optical centre coordinates from the NASA/IPAC Extragalactic Database (NED). 
\end{itemize}
\begin{table*}
        \centering
        \caption{3DB input parameters and characteristics of  data cubes.
        (1) Beam size of the data cube (NGC 2841, NGC 3198, and NGC 7331 are at the original resolution);
        (2) channel width;
        (3) noise in the channels;
        (4) noise in the total map;
        (5) outermost fitted radius;
        (6) sampling radius;
        (7,8) centre coordinate;
        (9) systemic velocity; and
        (10) normalisation method (L=local, A=azimuthal).}
        \label{tab:bb_params}
        \begin{tabular}{l|cccc|cccccc}
        \hline\hline
        Galaxy  & Beam          & $\Delta v$    & $\sigma_\mathrm{ch}$  & $\sigma_\mathrm{tot}$   & R$_\mathrm{max}$      & $\Delta$R     & centre RA      & centre DEC            & V$_\mathrm{sys}$      & norm          \\
                & ($\arcsec$)   & (\kms)        & \multicolumn{2}{c|}{(mJy/beam)}               & (kpc)                   & (pc)          & (hh mm ss.s)  & (dd mm ss)            & (\kms)          &       \\
                & (1)           & (2)           & (3)                   & (4)                     & (5)                   & (6)           & (7)           & (8)                     & (9)                   & (10)          \\
        \hline
        DDO 154 & 21 x 21       & 2.6           & 1.366                 & 6.3                     & 8.3                   & 412           & 12 54 6.35    & 27 09 0.5               & 375.2                 & L             \\
        IC 2574 & 22 x 22       & 2.6           & 1.793                 & 10                      & 9.6                   & 418           & 10 28 23.5    & 68 24 44                & 44.9                  & L             \\
        NGC 0925& 9 x 9         & 2.6           & 0.820                 & 5.002                   & 16.1                  & 401           & 02 27 16.9    & 33 34 45                & 546.3                 & L             \\
        NGC 2403& 27 x 27       & 5.2           & 0.0326                & 0.154                   & 19                    & 413           & 07 36 51.4    & 65 36 09                & 133.2                 & L             \\
        NGC 2841& 6.06 x 5.79   & 5.2           & 0.3917                & 2.745                   & 46.8                  & 410           & 09 22 02.6    & 50 58 35                & 633.7                 & A             \\
        NGC 2976& 23 x 23       & 5.2           & 0.934                 & 3.769                   & 3.2                   & 400           & 09 47 15.4    & 67 54 59                & 1.1                   & L             \\
        NGC 3198& 7.64 x 5.62   & 5.2           & 0.382                 & 0.198                   & 31.4                  & 474           & 10 19 54.9    & 45 32 59                & 657                   & A             \\
        NGC 4736& 18 x 18       & 5.2           & 0.895                 & 4.502                   & 7.8                   & 410           & 12 50 53.0    & 41 07 14                & 306.7                 & L             \\
        NGC 5055& 9 x 9         & 5.2           & 0.499                 & 2.839                   & 33.3                  & 432           & 13 15 49.3    & 42 01 45                & 496.7                 & L             \\
        NGC 6946& 15 x 15       & 2.6           & 1.556                 & 10.179          & 14.5                  & 402           & 20 34 52.3    & 60 09 14                & 48.7                  & L             \\
        NGC 7331& 4.94 x 4.6    & 5.2           & 0.505                 & 3.318                   & 24.8                  & 357           & 22 37 04.0    & 34 24 56                & 818.3                 & A             \\
        NGC 7793& 24 x 24       & 2.6           & 2.343                 & 16.634          & 7.14                  & 420           & 23 57 49.8    & -32 35 28               & 226.2                 & A             \\
        \hline
        \end{tabular}
\end{table*}

\section{Revisited mass model for NGC 7793}\label{ap:ngc7793}
\begin{figure}
    \includegraphics[width=1\columnwidth]{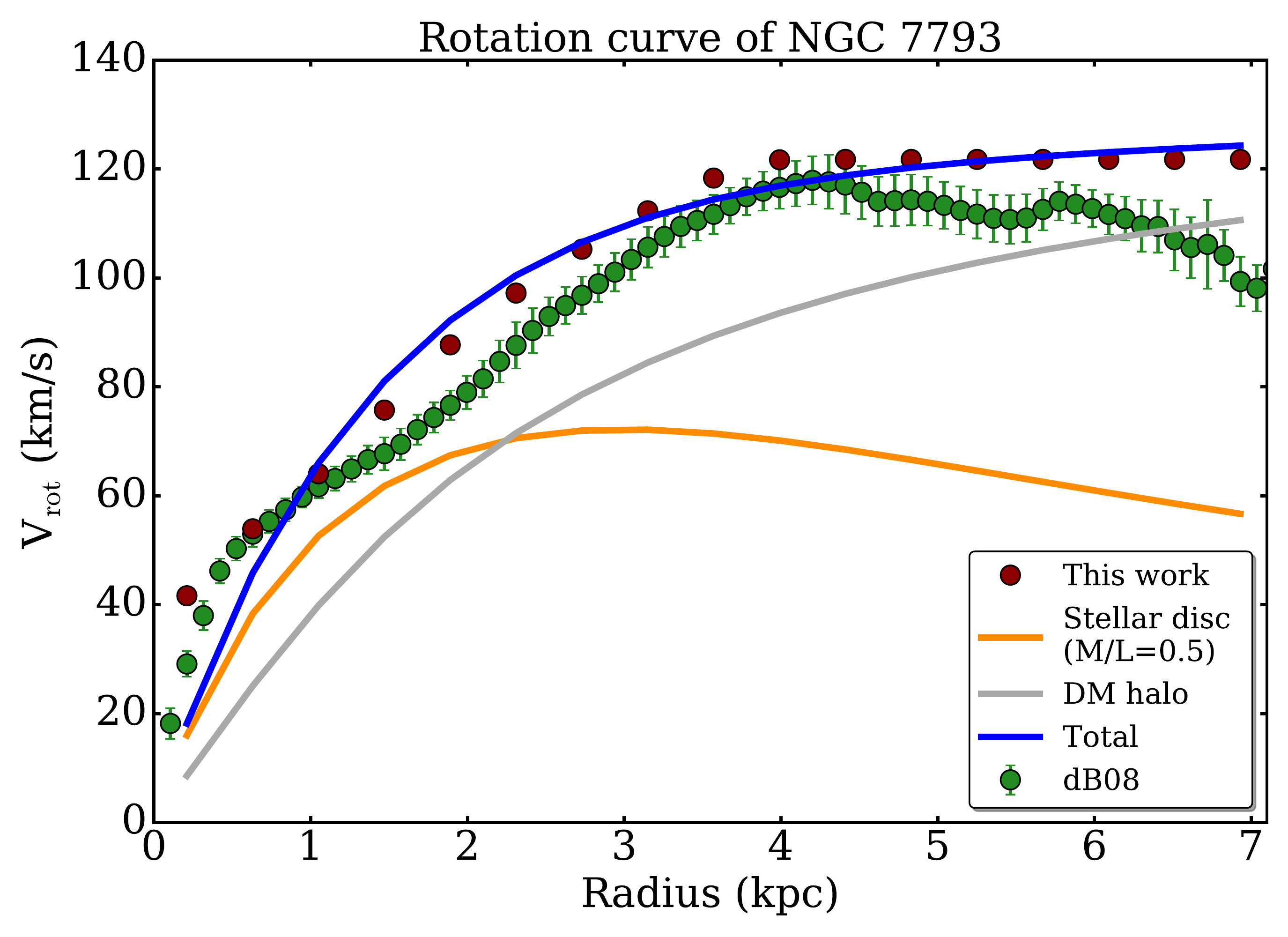}
    \caption{HI rotation curve of NGC 7793 assuming a flat rotation curve (red points), the errors are of the order of 4--5 \kms. 
    The solid lines show our mass model: the stellar disc with M/L=0.5 and the DM contributions are shown in orange and grey, while\ the resulting total 
    rotation curve is in blue. 
    The green points show the rotation curve measured by \citet{2008Deblok}.}
    \label{fig:vc_n7793}
\end{figure}

The observed HI rotation curve of NGC 7793 has a declining profile in \cite{2008Deblok} (green points in Fig.~\ref{fig:vc_n7793}). 
The authors interpreted this shape as a signature of a maximum stellar disc, but they found relatively low best-fit M/L values of 0.22 or 0.31 depending on the 
assumed initial mass function. 
In order to find a reliable mass model, we decided to interpret the declining rotation curve as the result of a small warp in inclination beyond 4 kpc. 
In practice, we first performed the tilted-ring fitting using 3DB to determine a first guess of the rotation curve. 
Then, we repeated the fit using this rotation curve for the rings at $R<4$ kpc but fixing V$_\mathrm{rot}$ at its peak (121.8 \kms) for the rings beyond 4 kpc. 
The resulting best-fit inclination starts to decrease at 4 kpc from about 44\textdegree\, to about 40\textdegree. 
In Fig.~\ref{fig:vc_n7793}, the difference between our rotation curve and that of \cite{2008Deblok} is due to the radial variation of the 
inclination. Our best-fit inclination is systematically lower than that of \cite{2008Deblok}, so our rotation curve tends to be higher.  
We note that both rotation curves are realistic, but removing the decreasing part allows us to find a parametric mass model for the DM halo that reproduces 
the observed rotation much better. 
However, line-of-sight warps are notoriously difficult to trace with fitting algorithms \citep{2003Gentile}. 
To set the mass model parameters, we fixed the M/L ratio of the stellar disc to 0.5 \citep{2016Lelli} and repeated the isothermal halo fit on the flat rotation curve, 
leaving $\rho_{\mathrm{DM},0}$ and $r_\mathrm{c}$ as free parameters. 
In Fig.~\ref{fig:vc_n7793}, we show that our model can reproduce the whole rotation curve. 
For completeness, we checked if the measurement of the velocity dispersion profile is influenced by the assumption of the flat rotation curve beyond 4 kpc. 
We found a slight offset between the $\sigma_\mathrm{HI}$ radial profiles with the flat and declining V$_\mathrm{rot}(R)$, but the two are fully compatible 
within the uncertainties.

\section{NGC 2841, an example of the scale height imprint on the velocity dispersion field}\label{ap:ngc2841_xshape}
\begin{figure}
\includegraphics[width=1.\columnwidth]{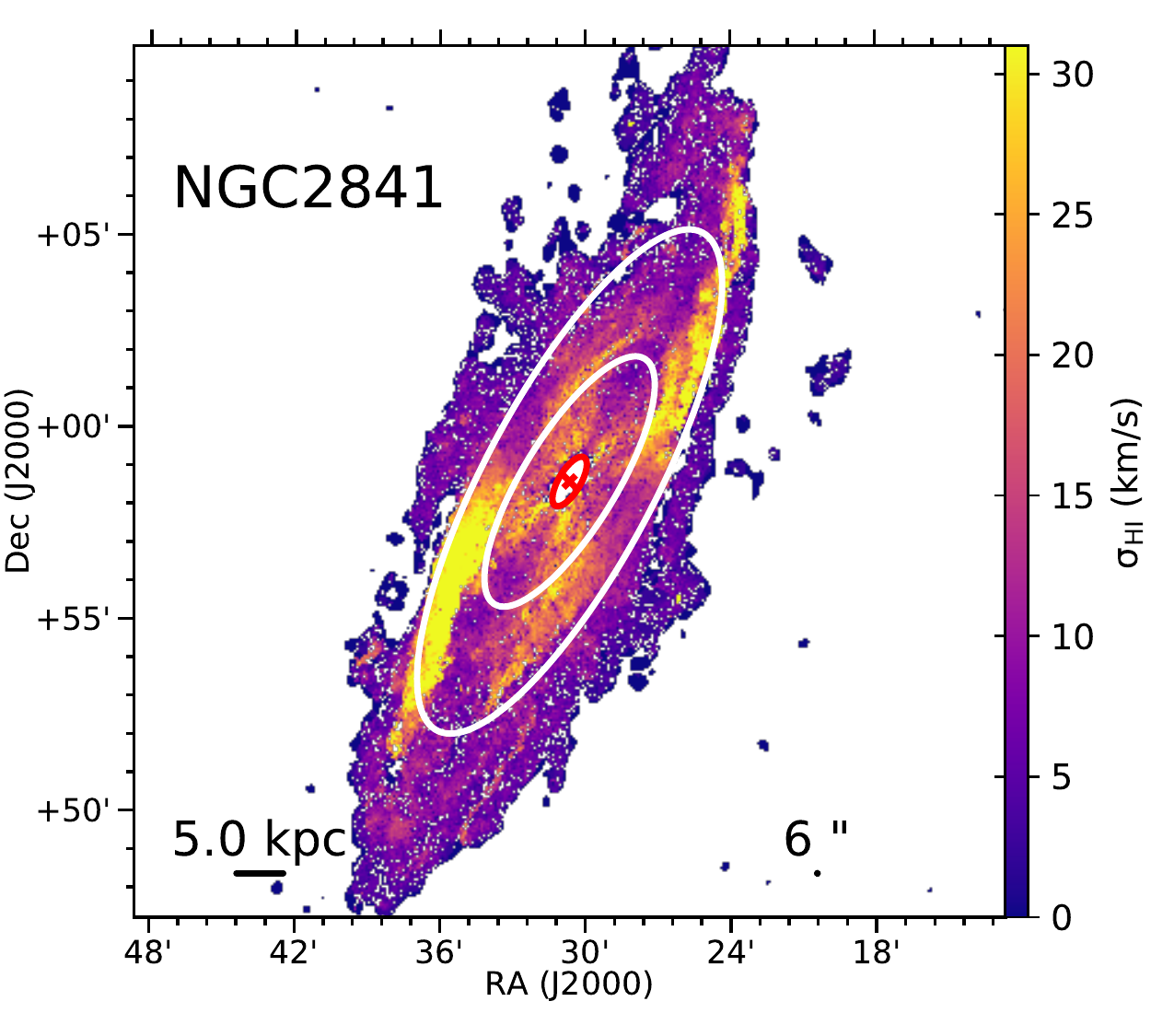}
\caption{HI velocity dispersion map of NGC 2841. We note the X-shaped region where $\sigma_{\mathrm{HI}}>$ 15 \kms; the white ellipses 
correspond to the $R=15$ kpc and $R=30$ kpc annuli. The red cross indicates the centre and the red ellipse delimits the gas depleted region. 
The black dot (lower right) shows the beam size.}
\label{fig:NGC2841_mom2_xshape}
\end{figure}

In Fig.~\ref{fig:vdisp_fits}, the velocity dispersion profile of  NGC 2841 shows a sudden increase by 10 \kms between 15 kpc to 30 kpc in radius. 
An inspection of the velocity dispersion map can help us understand the origin of this feature. 
In Fig.~\ref{fig:NGC2841_mom2_xshape}, the galaxy is coloured according to the velocity dispersion value in each pixel. 
The red cross shows the centre of the galaxy (Table~\ref{tab:bb_params}) and it is surrounded by an HI hole delimited by the red ellipse, which 
corresponds to an annulus of radius of 3 kpc. 
The centre of the galaxy is deficient in both HI and H$_2$, as also pointed out by \cite{2016Frank}. 
However, we want to focus on the most prominent feature of the map, which is the yellow X-shaped region with $\sigma_\mathrm{HI}>$15 \kms approximately 
delimited by the white ellipses. 
This X-shaped feature is typical of thick discs \citep{1997Sicking,2018Iorio} and warped galaxies, as it is due to different line-of-sight 
velocities being mixed in projections along selected directions. 
Therefore, the azimuthal average in these regions is biased towards high values, which is exactly what happens between the annuli with $R=15$ kpc and 
$R=30$ kpc. 
We conclude that the high velocity dispersion measured in this region is spurious and can be safely ignored in our modelling. 

The more the galaxy is inclined with respect to the line of sight, the more important is the effect of increasing the velocity dispersion. 
NGC 2841, NGC 3198, and NGC 7331 are the galaxies in the sample with $i>70$\textdegree.
The effect of inclination on $\sigma_\mathrm{HI}$ profile of NGC 3198 is less prominent with respect to NGC 2841, as the former galaxy is less inclined. 
On the other hand, the whole velocity dispersion profile of NGC 7331 is likely overestimated, but the associated uncertainties are large enough to account for 
this effect. 

\section{Uncertainties on scale heights and volume densities}\label{ap:error_prop}
In this section, we explain how the uncertainties on scale heights ($\Delta h$) and volume densities ($\Delta \rho$) are estimated. 
Let us first focus on the gas component and then on the SFR. 

For a gas disc in hydrostatic equilibrium, the scale height can be calculated using the approximated Eq.~\ref{eq:densprofscale}. 
Applying the rules of the propagation of uncertainty, we find
\begin{equation*}
 \Delta h = \left \{ 
            \left(h \frac{\Delta \sigma}{\sigma} \right)^2 + 
            \left[ 
            \frac{h}{2} \Delta \left( \frac{\partial^2 \Phi(R,0)}{\partial z^2} \right) 
            \left( \frac{\partial^2 \Phi(R,0)}{\partial z^2} \right)^{-1} 
            \right]^2
            \right \}^{\frac{1}{2}} \, ,
\end{equation*}
where $\Delta \sigma$ and $\Delta \left( \frac{\partial^2 \Phi(R,0)}{\partial z^2} \right)$ are the uncertainties on the velocity dispersion and the 
second derivative of the gravitational potential. 
In our case, $\Delta \sigma$ coincides with the error on the velocity dispersion measured by 3DB, but finding 
$\Delta \left( \frac{\partial^2 \Phi(R,0)}{\partial z^2} \right)$ would be much more problematic. 
The uncertainty on $\frac{\partial^2 \Phi(R,0)}{\partial z^2}$ is linked to the uncertainty on the total volume 
density of DM and stars through the Poisson equation, so we should use the uncertainties on the mass decomposition. 
However, we expect that the parametric mass models of stars and DM do not significantly change within the errors on the fit and the observed rotation 
curve in \cite{2008Deblok}. 
For simplicity, we assume that $\Delta \left( \frac{\partial^2 \Phi(R,0)}{\partial z^2} \right) \ll \Delta \sigma$, therefore
\begin{equation}
 \Delta h_\mathrm{HI} = h_\mathrm{HI} \frac{\Delta \sigma_\mathrm{HI}}{\sigma_\mathrm{HI}} \, ; \\
 \Delta h_\mathrm{H_2} = h_\mathrm{H_2} \frac{\Delta \sigma_\mathrm{H_2}}{\sigma_\mathrm{H_2}}
.\end{equation}
Then, the general equation for the uncertainties on the volume densities is derived from Eq.~\ref{eq:midplane}:
\begin{equation}\label{eq:error_rho}
 \Delta \rho = \rho \left[ \left( \frac{\Delta \Sigma}{\Sigma} \right)^2 +  \left( \frac{\Delta h}{h} \right)^2 \right]^{\frac{1}{2}} \, ,
\end{equation}
where $\Delta \Sigma$ is the uncertainty on the observed surface densities. 
Therefore, the errors on HI, H$_2$, and total gas volume densities are
\begin{equation}
\Delta \rho_\mathrm{HI} = \rho_\mathrm{HI} \left[ \left(\frac{\Delta \Sigma_\mathrm{HI}}{\Sigma_\mathrm{HI}} \right)^2 +  
\left( \frac{\Delta h_\mathrm{HI}}{h_\mathrm{HI}} \right)^2 \right]^{\frac{1}{2}} \, ,
\end{equation}
\begin{equation}
 \Delta \rho_\mathrm{H_2} = \rho_\mathrm{H_2} \left[ \left( \frac{\Delta \Sigma_\mathrm{H_2}}{\Sigma_\mathrm{H_2}} \right)^2 +  
\left( \frac{\Delta h_\mathrm{H_2}}{h_\mathrm{H_2}} \right)^2 \right]^{\frac{1}{2}} \, ,
\end{equation}
\begin{equation}
 \Delta \rho_\mathrm{gas} = \left ( \Delta \rho_\mathrm{HI}^2 + \Delta \rho_\mathrm{H_2}^2 \right )^{\frac{1}{2}} \, .
\end{equation}
We neglect the covariance between the error on $\Sigma_\mathrm{HI}$ and on $h_\mathrm{HI}$ as the scale height depends on the dominant mass components (the stellar 
disc and DM halo), so these two quantities can be considered independent; the same is valid for the molecular gas.

Concerning the constant SFR scale height, the uncertainty is null by construction ($\Delta h_\mathrm{SFR} = 0$), so the error on the SFR volume 
density derived from Eq.~\ref{eq:error_rho} is
\begin{equation}
 \Delta \rho_\mathrm{SFR} = \frac{ \Delta \Sigma_\mathrm{SFR} }{ \sqrt{2 \pi} h_\mathrm{SFR} } \, .
\end{equation}
In the case of the flaring SFR scale height, the error is derived from Eq.~\ref{eq:hSFR_wmean}
\begin{equation}\label{eq:error_hsfr}
\begin{split}
 \Delta h_\mathrm{SFR} & = 
 \Biggl[
  \left ( \frac{\Sigma_\mathrm{HI} \Delta h_\mathrm{HI}}{\Sigma_\mathrm{gas}} \right )^2+
  \left ( \frac{\Sigma_\mathrm{H_2} \Delta h_\mathrm{H_2}}{\Sigma_\mathrm{gas}} \right )^2+ \\
  & + \frac{ \left ( \Sigma_\mathrm{H_2}^2 \Delta \Sigma_\mathrm{HI}^2 + \Sigma_\mathrm{HI}^2 \Delta \Sigma_\mathrm{H_2}^2 \right )
  \left ( h_\mathrm{HI} - h_\mathrm{H_2} \right )^2}
  { \Sigma_\mathrm{gas}^4}
  \Biggr]^{\frac{1}{2}} \, .
\end{split}
\end{equation}
Therefore, the uncertainty on the volume density is
\begin{equation}\label{eq:error_rhosfr_gas}
 \Delta \rho_\mathrm{SFR} = \rho_\mathrm{SFR} \left[ \left(\frac{\Delta \Sigma_\mathrm{SFR}}{\Sigma_\mathrm{SFR}} \right)^2 +  
\left(\frac{\Delta h_\mathrm{SFR}}{h_\mathrm{SFR}} \right)^2 \right]^{\frac{1}{2}} \, .
\end{equation}

When we build the HI-SFR and H$_2$-SFR relations, we are implicitly assuming that  $h_\mathrm{SFR} = h_\mathrm{HI}$ and $h_\mathrm{SFR} = h_\mathrm{H_2}$, 
so  $\Delta h_\mathrm{SFR} = \Delta h_\mathrm{HI}$ and $ \Delta h_\mathrm{SFR} = \Delta h_\mathrm{H_2}$. 
As a consequence, Eq.~\ref{eq:error_rhosfr_gas} becomes
\begin{equation}
 \Delta \rho_\mathrm{SFR} =\rho_\mathrm{SFR} \left[ \left( \frac{\Delta \Sigma_\mathrm{SFR}}{\Sigma_\mathrm{SFR}} \right)^2 +  
\left( \frac{\Delta h_\mathrm{HI}}{h_\mathrm{HI}} \right)^2 \right]^{\frac{1}{2}} \, ,
\end{equation}
\begin{equation}
\Delta \rho_\mathrm{SFR} = \rho_\mathrm{SFR} \left[ \left( \frac{\Delta \Sigma_\mathrm{SFR}}{\Sigma_\mathrm{SFR}} \right)^2 +  
\left(\frac{\Delta h_\mathrm{H_2}}{h_\mathrm{H_2}} \right)^2 \right]^{\frac{1}{2}} \, .
\end{equation}

\section{Likelihood and posterior distributions of Bayesian fittings}\label{ap:posteriors}
In order to include both the orthogonal intrinsic scatter and the $x$ and $y$ errors on the volume densities, the logarithmic likelihood is written as 
\citep{2017Ponomareva,2018Posti} 
\begin{equation}\label{eq:loglike}
 \ln \mathcal{L} = - \frac{1}{2} \sum_{i=1}^N \left[ \frac{d_i^2}{\sigma_\mathrm{tot}^2} + \ln (2 \pi \sigma_\mathrm{tot}^2) \right ] \, .
\end{equation}
In this equation, $N$ is the number of data points, $d_i$ is the distance between a given data point $(x_i, y_i)$ and the model (e.g. Eq.~\ref{eq:vsf_law_log}), 
being $x_i = \log \rho_\mathrm{gas}$ and $y_i =  \log \rho_\mathrm{SFR}$. 
Then, $\sigma_\mathrm{tot}^2 = \sigma_\perp^2 + \sigma_{x_i,\perp}^2 + \sigma_{y_i,\perp}^2$, where $\sigma_\perp$ is the orthogonal intrinsic scatter and 
$\sigma_{x_i,\perp}= \sigma_{x_i} \cos \theta$ and $\sigma_{y_i,\perp}= \sigma_{y_i} \sin \theta$ are the projections of the $x$ and $y$ uncertainties on 
data points $\sigma_{x_i}$ and $\sigma_{y_i}$ using the angle $\theta$, which is the arctangent of the slope of the relation. 
The prior distribution of the free parameters is uniform and spans from $-\infty$ to $+\infty$ for the slope and $y$-intercept, and from 0 to 
$+\infty$ for the intrinsic scatter. 
Before the fitting, the origin of the data points coordinate system is shifted to the median of $\rho_\mathrm{gas}$ and $\rho_\mathrm{SFR}$ 
($x_\mathrm{m}$ and $y_\mathrm{m}$ in logarithmic scale) to reduce the covariance between $\alpha$ and $\log A$. 
In practice, the axes of the new coordinate system $x'-y'$ are defined as $y'=y- y_\mathrm{m}$ and $x'=x- x_\mathrm{m}$, therefore the best-fit $\log A$ in the 
$x-y$ system becomes $\log A = y_\mathrm{m} - \alpha x_\mathrm{m} + \log A'$, where $\log A'$ is the best-fit intercept in the $x'-y'$ system. 
Clearly, when we consider only the atomic (molecular) gas phase in Sec.~\ref{sec:h2only_corr} (Sec.~\ref{sec:hionly_corr}), we use the same 
method but with $x_i = \log \rho_\mathrm{HI}$ ($x_i = \log \rho_\mathrm{H_2}$) and the slope and normalisation are defined as $\beta$ and $\log B$ 
($\gamma$ and $\log \Gamma$). 
In the case with total gas, $x_\mathrm{m}=-1.94$ and $y_\mathrm{m}=-2.30$ ($y_\mathrm{m}=-2.74$) with the constant (flaring) $h_\mathrm{SFR}$. 
In the case with atomic gas, $x_\mathrm{m}=-2.02$ and $y_\mathrm{m}=-2.30$ ($y_\mathrm{m}=-2.76$) with the constant (flaring) $h_\mathrm{SFR}$. 
In the case with molecular gas, $x_\mathrm{m}=-2.05$ and $y_\mathrm{m}=-1.86$. 

\begin{figure*}
\centering
\subfloat[][\label{fig:triangle_hsfr_const_xpaper}]
{\includegraphics[width=.335\textwidth]{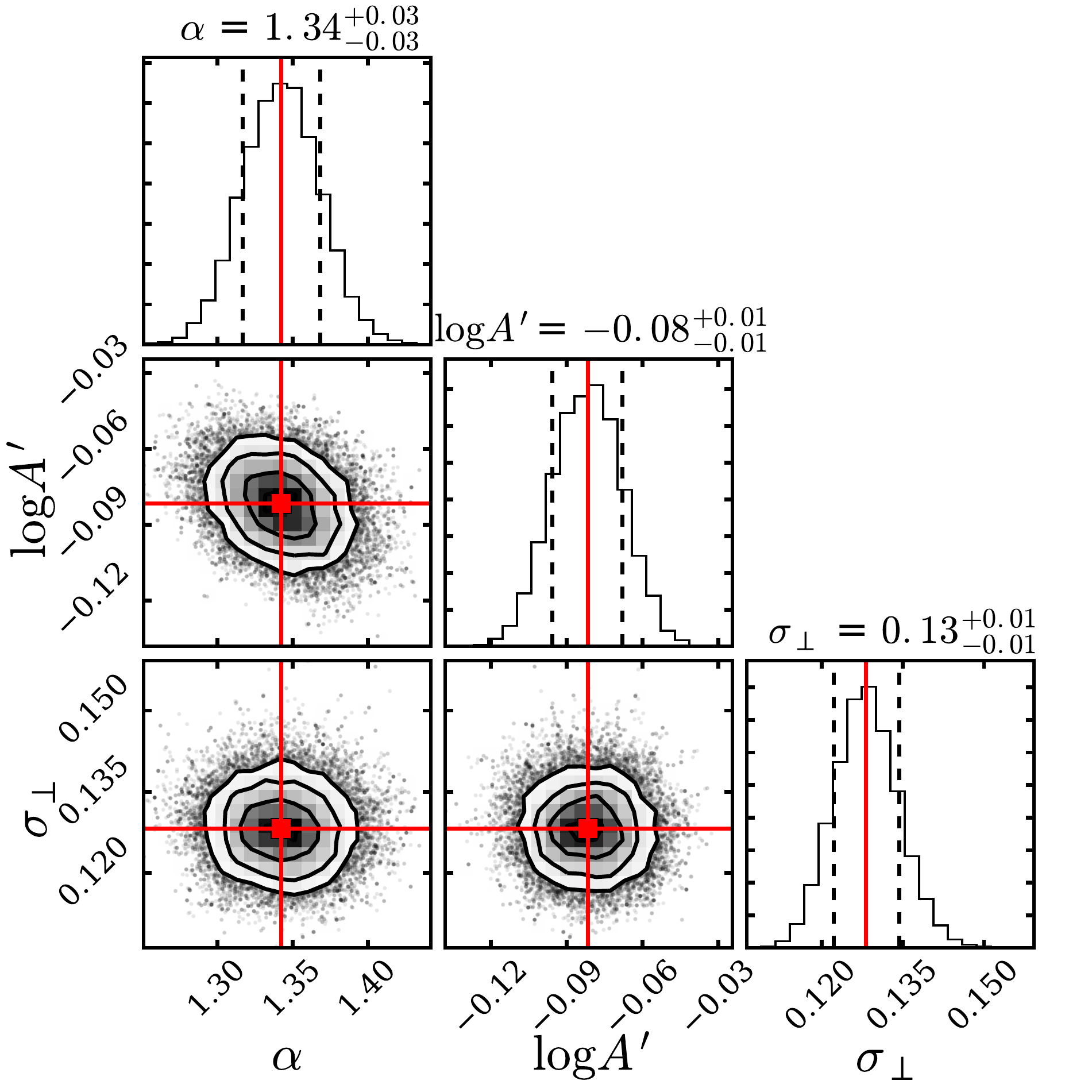}}
\subfloat[][\label{fig:triangle_hsfr_const_xpaper_hi}]
{\includegraphics[width=.335\textwidth]{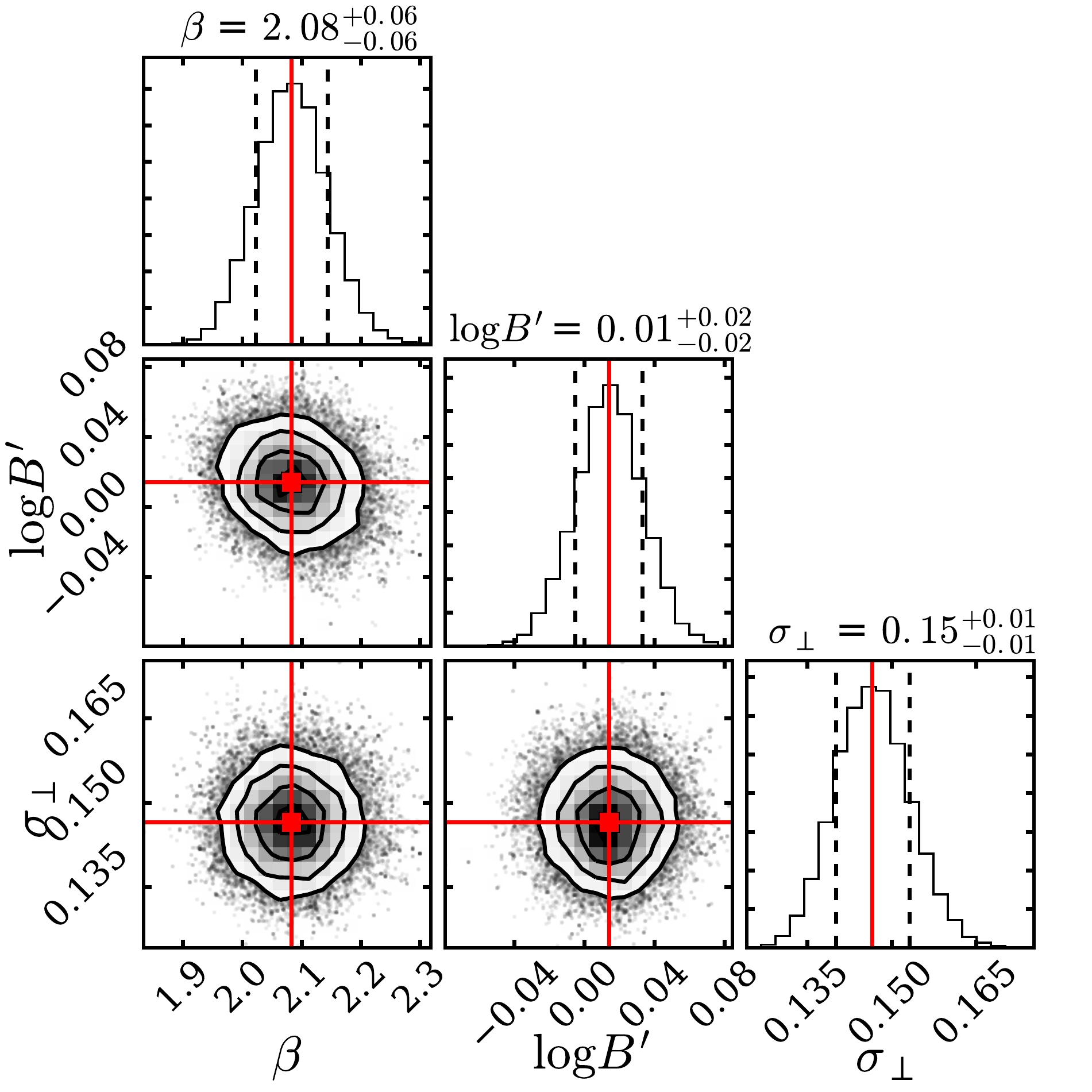}}
\subfloat[][\label{fig:triangle_hsfr_const_xpaper_h2}]
{\includegraphics[width=.335\textwidth]{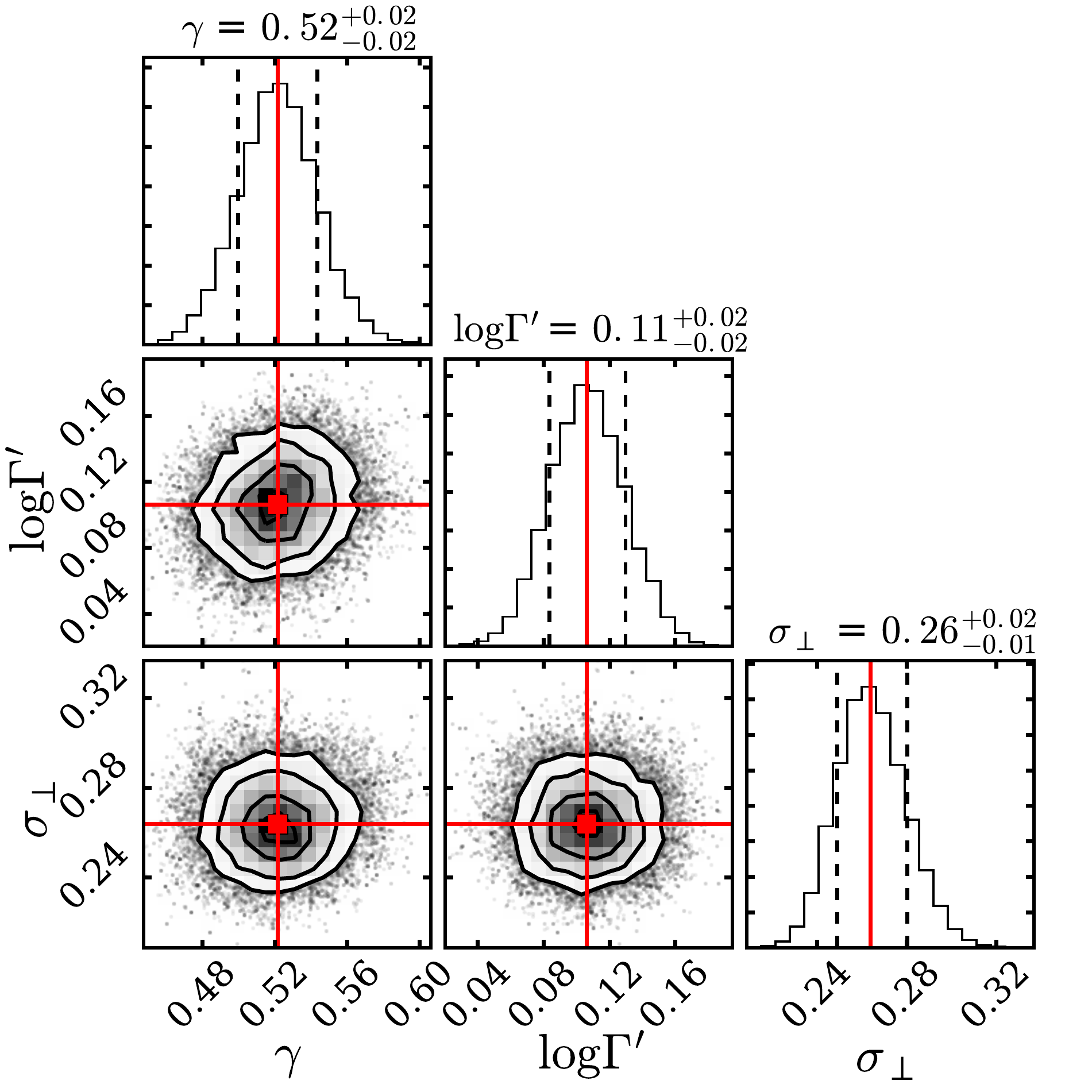}}
\\
\subfloat[][\label{fig:triangle_hsfr_flare_xpaper}]
{\includegraphics[width=.335\textwidth]{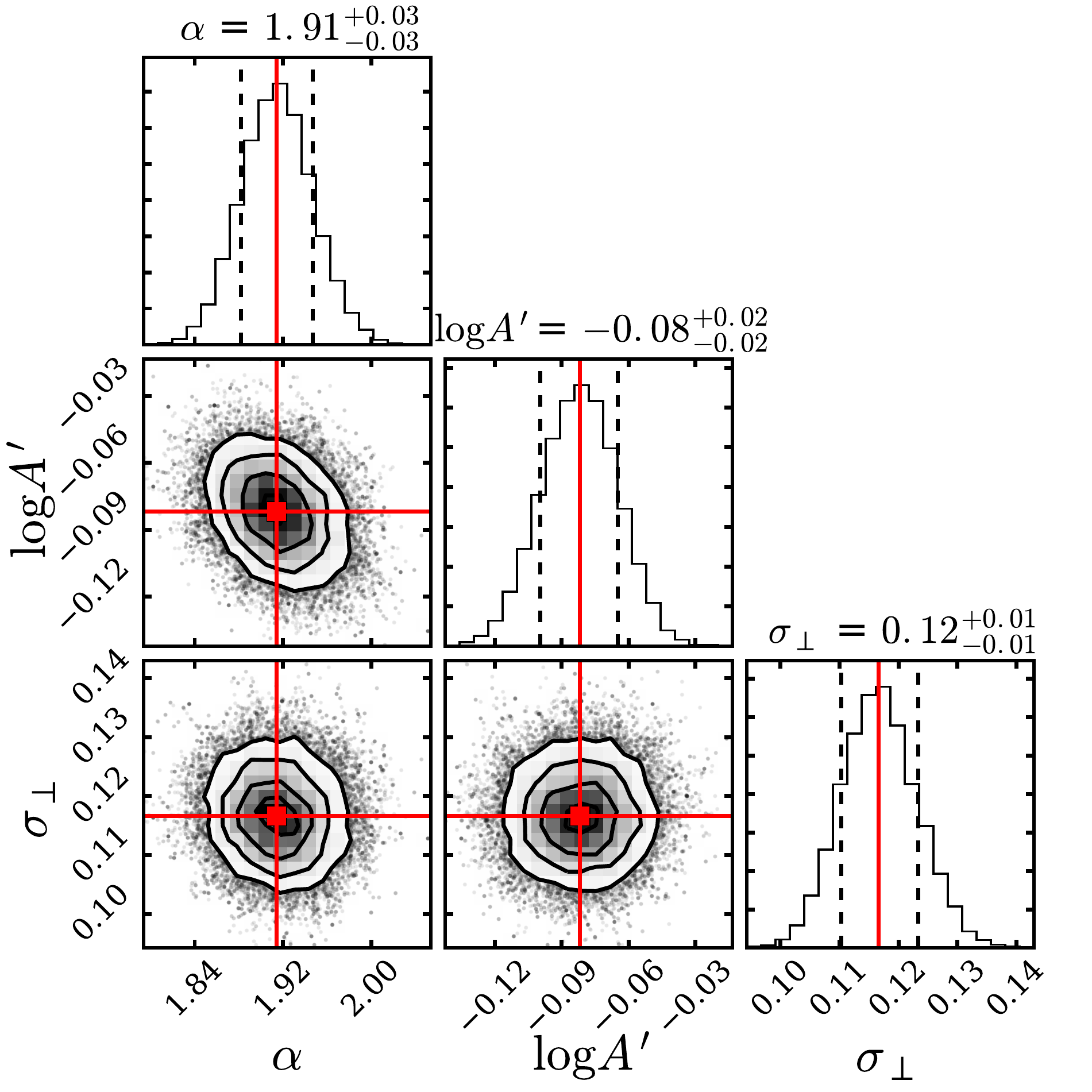}}
\subfloat[][\label{fig:triangle_hsfr_flare_xpaper_hi}]
{\includegraphics[width=.335\textwidth]{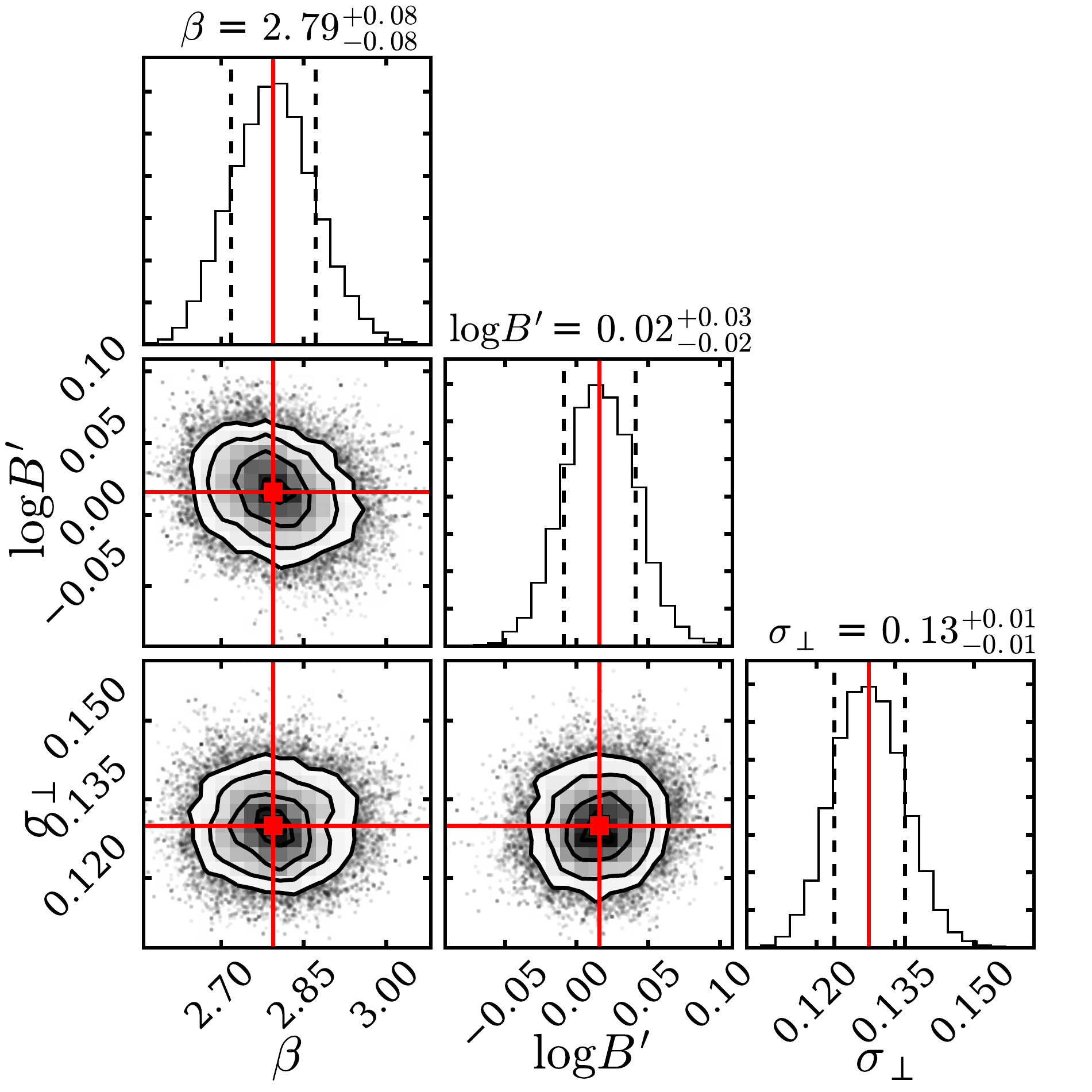}}
\subfloat[][\label{fig:triangle_hsfr_flare_xpaper_h2}]
{\includegraphics[width=.335\textwidth]{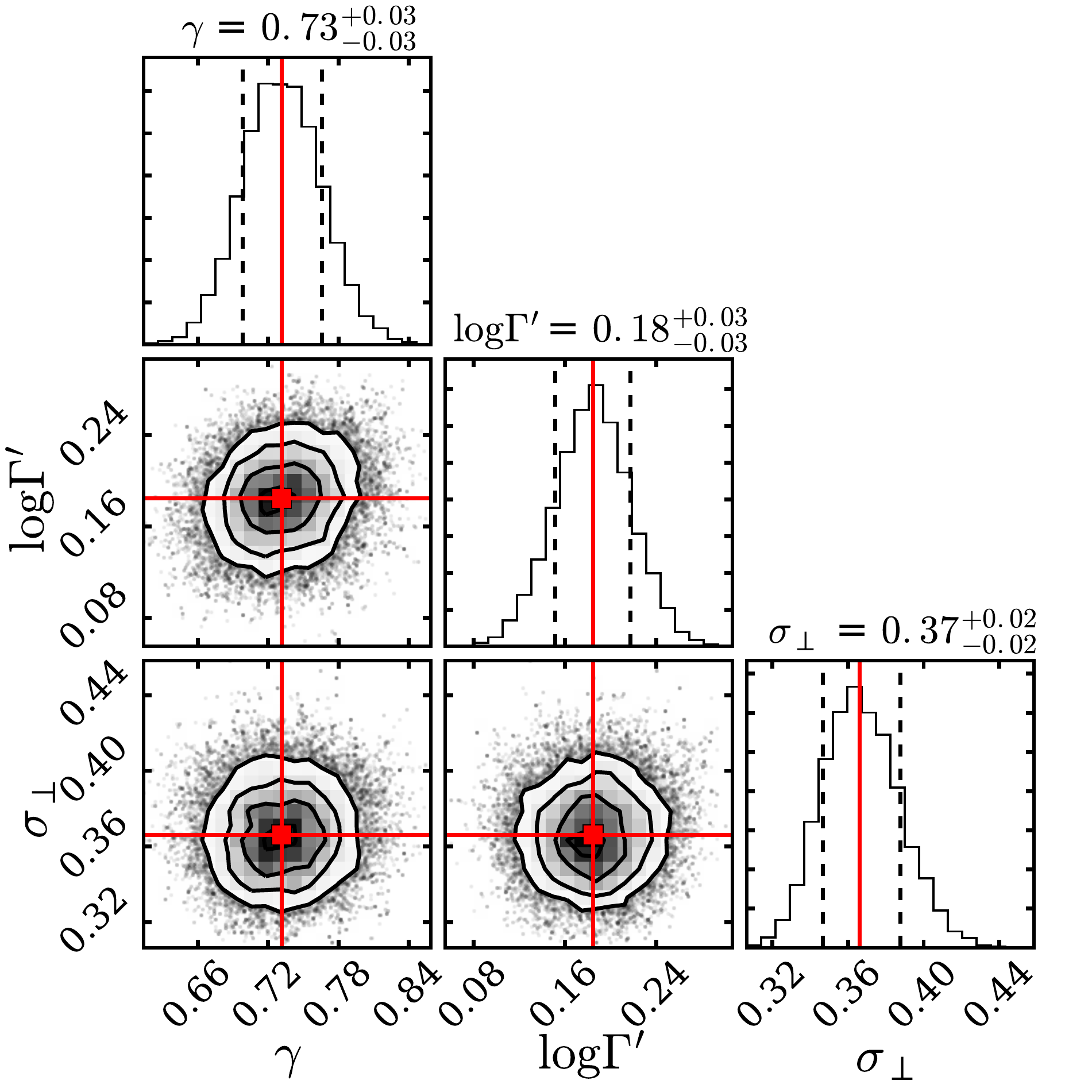}}
\caption{Marginalised posterior distributions of free parameters in the Bayesian fittings of VSF laws with total gas (left column), HI only (central column), 
and H$_2$ only (right column). In the first row $h_\mathrm{SFR}$ is constant at 100 pc, while in the second row it flares with radius (Eq.~\ref{eq:hSFR_wmean}). 
The contours of the 2D posteriors (corners) encompass the 68.3\%, 86.6\%, and 95.4\% probability and the red squares and lines indicate the best-fit 
parameters. In the 1D posteriors (diagonals), the best-fit parameters are indicated by the red line and the 16th and 84th percentiles are indicated by the 
black dashed lines.}
\label{fig:corner_gas}
\end{figure*}

Fig.~\ref{fig:corner_gas} shows the marginalised posterior distributions of free parameters for the VSF laws MCMC fittings. 
The first column refers to the VSF law between the SFR and total gas (Eq.~\ref{eq:vsf_law_log}). 
Despite the axes shifting, there is still a small covariance between $\alpha$ and $\log A'$ but all the parameters are well constrained and clearly indicate 
the existence of a correlation between the volume densities of gas and SFR. 
Unfortunately, we are not able to find an unambiguous best-fit slope, as it depends on the choice of $h_\mathrm{SFR}$, but the intrinsic scatter is small 
(0.12-0.13 dex) in both cases. 
The second and third columns are the same as the first but for the VSF laws with HI (Eq.~\ref{eq:vsf_law_log_hi}) and H$_2$ 
(Eq.~\ref{eq:vsf_law_log_h2}).  
The relation with HI is steeper and has a small scatter compared to the relation with H$_2$. 
%

\end{document}